\documentclass[journal,twocolumn]{IEEEtran}
\newcommand{\mysize}{0.48}
\usepackage{xcolor}
\usepackage{array}
\usepackage{generic}
\usepackage{textcomp}
\usepackage{cite}
\usepackage{amsmath,amssymb,amsfonts}
\usepackage{algorithmic}
\usepackage{graphicx}
\usepackage{mathrsfs}
\usepackage{caption}
\usepackage{subcaption}
\usepackage{soul}
\usepackage{array}
\newcommand{\varA}[1]{{\operatorname{#1}}}
\newcolumntype{P}[1]{>{\centering\arraybackslash}p{#1}}
\newcolumntype{M}[1]{>{\centering\arraybackslash}m{#1}}

\begin{document}

\title{Data-driven predictive control with improved performance using segmented trajectories}
\author{E. O’Dwyer*, E. C. Kerrigan, P. Falugi, M. A. Zagorowska and N. Shah
\thanks{* Corresponding author.}
\thanks{E. O'Dwyer is with the Department of Chemical Engineering, Imperial College London, email: e.odwyer@imperial.ac.uk}
\thanks{P. Falugi is with the Department of Electrical and Electronic Engineering, Imperial College London, email: p.falugi@imperial.ac.uk}
\thanks{E. C. Kerrigan is with the Department of Electrical and Electronic Engineering, Department of Aeronautics, Imperial College London, email: e.kerrigan@imperial.ac.uk}
\thanks{M. A. Zagorowska was with the Department of Electrical and Electronic Engineering, Imperial College London, email: m.zagorowska@imperial.ac.uk, currently with Automatic Control Laboratory, ETH Zurich, email:mzagorowska@control.ee.ethz.ch}
\thanks{N. Shah is with the Department of Chemical Engineering, Imperial College London, email: n.shah@imperial.ac.uk}
\thanks{For the purpose of open access, the author has applied a Creative Commons Attribution (CC BY) license to any Author Accepted Manuscript version arising.}
}

\maketitle

\begin{abstract}
A class of data-driven control methods has recently emerged based on Willems' fundamental lemma. Such methods can ease the modelling burden in control design but can be sensitive to disturbances acting on the system under control. In this paper, we propose a restructuring of the problem to incorporate segmented prediction trajectories. The proposed segmentation leads to reduced tracking error for longer prediction horizons in the presence of unmeasured disturbance and noise when compared to an unsegmented formulation. The performance characteristics are illustrated in a set-point tracking case study in which the segmented formulation enables more consistent performance over a wide range of prediction horizons. The method is then applied to a building energy management problem using a detailed simulation environment. The case studies show that good tracking performance is achieved for a range of horizon choices, whereas performance degrades with longer horizons without segmentation.

\end{abstract}


\section{Introduction}

\IEEEPARstart{T}{he} increased focus on digital technology in recent times has drawn attention to data-driven control methods, with applications ranging from building control~\cite{Maddalena2020a}, to autonomous vehicles~\cite{Bhattacharyya2021}. By reducing the modelling burden in the control design phase, the deployment of advanced control can be streamlined, with control decisions directly obtained from data measurements~\cite{Wang2019b}. Control methods that rely directly on measurements, without using explicit models, are called \emph{direct data-driven methods}. In data-rich environments, such methods may then be advantageous. Nonetheless, direct data-driven methods often lack the theoretical foundations of \emph{indirect data-driven methods}, whereby data is needed to explicitly derive a model which is then used for control~\cite{DePersis2020}.

To overcome this, Willems' fundamental lemma~\cite{Willems2005}, has recently been used as a foundation for a large body of data-driven control research, because it provides theoretical foundations for control decisions obtained directly from data. Methods stemming from the lemma rely on the insight that, under certain straightforward excitation conditions, historical data structures can be used to project all permissible trajectories of a Linear Time-Invariant (LTI) system. By building on the fundamental lemma, theoretical foundations can be achieved for certain direct data-driven methods. Predictive control representations have been developed that can offer stability and certain robustness guarantees~\cite{DePersis2020}, without requiring the derivation of a parametric model. For example, a data-driven counterpart to Model Predictive Control (MPC) named data-enabled predictive control (DeePC) was proposed in~\cite{Coulson2019a} and shown to be competitive with MPC in the deterministic case. In~\cite{Fiedler2020}, an equivalence was then shown between this direct data-driven approach and an alternative subspace-based approach (Subspace Predictive Control, SPC), in which the parameters of a multi-step prediction model are derived using the same training data criteria. This was expanded upon in~\cite{Dorfler2021}, where further analysis of the performance of these direct and indirect formulations was carried out for different systems using various relaxation and regularisation techniques. The results suggested that noisy data have a greater impact on direct formulations, while system nonlinearities have a greater impact on indirect formulations. Additionally, rather than relying on a single training period for data acquisition, a strategy was developed in~\cite{VanWaarde2020} by which multiple, potentially short, datasets can be used instead, without compromising the theoretical foundation of the fundamental lemma.

Uncertainty in data measurements will impact the performance of a data-driven approach, thus several methods have been developed to ensure viability in stochastic settings. In~\cite{Alpago2020}, the authors supplement a data-driven controller with a data-driven extended Kalman filter to reduce sensitivity to noise. Robust formulations have also been developed to enable performance guarantees under certain conditions of system stochasticity, such as the robust modification proposed in~\cite{Berberich2021}, ensuring exponential stability in the presence of measurement noise. In~\cite{Coulson2020}, a chance-constrained distributionally robust formulation was developed for stochastic linear, time-invariant (LTI) systems, providing probabilistic guarantees on performance. Tractable, robust formulations are proposed to ensure performance guarantees under uncertainty in~\cite{Huang2021}, while in~\cite{Lian2021}, a correspondence was found between the fundamental lemma perspective and that of System-Level Synthesis (SLS), which was then exploited to formulate a robust closed-loop data-predictive controller. A robust building-level controller was implemented in a real building in \cite{Lian2021b} whereby an active excitation method was used to allow for the data structures to be updated continually while maintaining persistence of excitation.

In the presence of noise or unmeasured disturbances, data-driven approaches use regularisation and relaxation to attain robustness guarantees and to align the DeePC problem to more established Subspace Predictive Control (SPC) methods~\cite{Dorfler2021}. However, while methods exist to preclude acausal relationships between inputs and outputs in SPC, the lack of an explicit characterisation of the input-output relationship in direct DeePC mean that enforcing such conditions is not straightforward. Performance under noise and disturbance can then be affected, particularly for longer prediction horizons. In this paper, we propose a restructuring of the data-enabled predictive controller formulation whereby the prediction trajectory is divided into multiple shorter trajectories, denoted \emph{segments}. These segments can be identified in the same manner as the unsegmented formulation, with less training data. By exploiting the problem structure, computational effort can be reduced for longer horizon choices. This restructuring precludes most potential acausal dependencies between inputs and outputs, which could otherwise lead to degraded performance. The method is analysed and compared to the unsegmented version in a set-point tracking case study with different horizon lengths and disturbance and noise realisations. With segmentation, a performance improvement is shown in terms of set-point tracking error, while a linear computational time increase is observed for increasing horizon lengths, which favourably compares to the nonlinear increase observed for an unsegmented formulation. Following this, a building energy management case study is implemented, based on a detailed building simulation environment with realistic disturbances. Comfort and energy cost objectives are solved in a prioritised manner. As the segmented-trajectory approach behaves more consistently than the unsegmented approach for longer horizon lengths, a reduction in both cost and energy consumption is achieved for a one-day-ahead prediction horizon.

In Section~\ref{sec:II}, a background to the unsegmented data-driven predictive approach is provided based on the fundamental lemma, followed by the proposed modifications that result in a segmented formulation. In Section~\ref{sec:example1}, a set-point tracking case study is presented, with an analysis provided of the control performance and computational time associated with the segmented and unsegmented formulations. This is followed in Section~\ref{sec_bldg} by the building energy management case study, which is used to illustrate the benefits of the segmented formulation in a relevant application. The paper ends with conclusions in Section \ref{sec:End}.

\section{Modified data-driven predictive control formulation}\label{sec:II}
\subsection{Data-driven predictive control preliminaries}\label{sec:IIA}
As noted in the introduction, different variations of data-predictive control have been proposed. We provide a brief overview of the direct data-enabled predictive controller of~\cite{Coulson2019a} and the indirect multi-step prediction approach of~\cite{Fiedler2020} and~\cite{Dorfler2021} in this section.

A discrete-time~$n^{\text{th}}$-order LTI state-space system can be represented at sample instant $k$ by: 
\begin{equation}\label{eq:SSmod}
    \begin{aligned}
        x[k+1]=Ax[k]+Bu[k]\\
        y[k]=Cx[k]+Du[k],
    \end{aligned}
\end{equation}
where $x[k]\in\mathbb{R}^{n}$ is the system state-vector, $u[k]\in\mathbb{R}^{m}$ and $y[k]\in\mathbb{R}^{p}$ are the input and output vectors, respectively, and $A\in\mathbb{R}^{n\times n}$, $B\in\mathbb{R}^{n\times m}$, $C\in\mathbb{R}^{p\times n}$ and $D\in\mathbb{R}^{p\times m}$ are parameter matrices, the parameters of which are assumed to be unknown. Following the terminology of~\cite{Willems2007}, we define~$\mathscr{B}$ as the \textit{behaviour} of (\ref{eq:SSmod}), where the behaviour is defined as the set of possible outcomes of the system. The lag of the system is denoted~$\ell$, defined as the smallest integer for which the observability matrix $\mathscr{O}_{\ell}(A,C):=\left[C,CA,\ldots,CA^{\ell-1}\right]$ has full rank. By Willems' fundamental lemma~\cite{Willems2005}, arbitrary input and output sequences can be derived from a sufficiently long set of input/output data without explicitly estimating the parameters of (\ref{eq:SSmod}). The input-output sequences will be called \emph{trajectories}.
A representation of the system can then be found and used for predictive control only in terms of measured data. 

An offline data collection procedure is carried out to achieve this in which $T_{0}\in\mathbb{Z}_{>0}$ sequences of \textit{persistently exciting} input and output data measurements are given as $u_{tr}=\left[u^{T}_{1},\ldots,u^{T}_{T_{0}}\right]^{T}\in\mathbb{R}^{mT_{0}}$ and $y_{tr}=\left[y^{T}_{1},\ldots,y^{T}_{T_{0}}\right]^{T}\in\mathbb{R}^{pT_{0}}$ respectively with $\mathbb{Z}_{>0}$ denoting the set of positive integers. A trajectory $w$ is defined as persistently exciting of order $L_0$, $L_0\in\mathbb{Z}_{>0}$, if the Hankel matrix~$\mathscr{H}_{L_{0}}(w)$ is of full row-rank with
\begin{equation}\label{eq:Hank}
    \mathscr{H}_{L_{0}}(w):=\left[\begin{matrix}w_{1} & \cdots & w_{T_{0}-L_{0}+1}\\\vdots &\ddots &\vdots \\w_{L_{0}} & \cdots & w_{T_{0}}\end{matrix}\right].
\end{equation}
Note that non-square Hankel matrices are permitted in this definition. From~\cite{Willems2005}, for a controllable, observable~$\mathscr{B}$, if $w\in\mathscr{B}$ is a persistently exciting, $T_{0}$-samples-long trajectory of order~$t+n$, then any $t$-samples long trajectory in~$\mathscr{B}$ can be described as a linear combination of the columns of~$\mathscr{H}_{t}(w)$, and any~$\mathscr{H}_{t}(w)g$ is a trajectory of~$\mathscr{B}$ where~$g\in\mathbb{R}^{T-t+1}$. For persistent excitation, $T_{0}\geq(m+1)(t+n)-1$. Here we seek to construct trajectories of length~$N+T_{ini}$, where $N\in\mathbb{Z}_{>0}$ is the prediction horizon and $T_{ini}\in\mathbb{Z}$ is some initialisation length. Following~\cite[Lem.~1]{Markovsky2008}, by fixing the first~$T_{ini}$ samples of a trajectory, the subsequent $N$ samples are uniquely specified if~$T_{ini}\geq\ell$. 

The training data sequences~$u_{tr}$ and~$y_{tr}$ are arranged in the Hankel form of (\ref{eq:Hank}) with $T_{0}\geq(m+1)(T_{ini}+N+n)-1$ and $L_{0}=T_{ini}+N$.
The training data structures can then be defined at this point as $U_{tr}:=\mathscr{H}_{T_{ini}+N}(u_{tr})$ and $Y_{tr}:=\mathscr{H}_{T_{ini}+N}(y_{tr})$. These matrices are then partitioned such that the first~$T_{ini}$ block rows of~$U_{tr}$ and~$Y_{tr}$ are denoted by the subscript~$\alpha$ and are referred to as initialisation data, with the remaining rows denoted by~$\beta$ and referred to as prediction data. The partitioned data matrices are thus defined as
\begin{equation}
    \begin{aligned}
        \left[\begin{matrix}U_{\alpha}\\U_{\beta}\end{matrix}\right]:=\mathscr{H}_{T_{ini}+N}(u_{tr}),\\
        \left[\begin{matrix}Y_{\alpha}\\Y_{\beta}\end{matrix}\right]:=\mathscr{H}_{T_{ini}+N}(y_{tr}).
    \end{aligned}
\end{equation}

Defining initialisation sequences~$u_{ini}\in\mathbb{R}^{mT_{ini}}$ and $y_{ini}\in\mathbb{R}^{pT_{ini}}$ as the~$T_{ini}$ most recent measurements, any future trajectories~$u_{f}\in\mathbb{R}^{mN}$ and~$y_{f}\in\mathbb{R}^{pN}$ can be found as the solution to
\begin{equation}\label{eq:dirg0}
    \left[\begin{matrix}
    U_\alpha\\Y_\alpha\\U_\beta\\Y_\beta
    \end{matrix}\right]g=\left[\begin{matrix}
    u_{ini}\\y_{ini}\\u_{f}\\y_{f}
    \end{matrix}\right],
\end{equation}
where~$g\in\mathbb{R}^{T_{0}-T_{ini}-N+1}$. In this form, the data structures on the left-hand side of (\ref{eq:dirg0}) contain the persistently excited training data, whereas the right-hand side contains the predicted trajectories, divided into \textit{initial}, $u_{ini}$, $y_{ini}$, and \textit{future}, $u_{f}$, $y_{f}$, portions.

This leads to the insight that~$u_{f}$ and~$y_{f}$, the future trajectories of~$\mathscr{B}$, can be found for a given training data set and given initialisation trajectories~$u_{ini}$ and~$y_{ini}$. From this, a Data-enabled Predictive Control (DeePC) formulation was proposed in~\cite{Coulson2019a}, whereby the following optimisation is carried out:
\begin{equation}\label{eq:dir1}
    \min_{g,u_{f},y_{f}}V\left(g,u_{f},y_{f}\right)
\end{equation}
$\qquad\qquad\text{s.t.}$
\begin{eqnarray}
    \left[\begin{matrix}
    U_\alpha\\Y_\alpha\\U_\beta\\Y_\beta
    \end{matrix}\right]g &=&\left[\begin{matrix}
    u_{ini}\\y_{ini}\\u_{f}\\y_{f}
    \end{matrix}\right]\label{eq:dirg}\\
    u_{f} &\in&\mathcal{U}\\
    y_{f} &\in&\mathcal{Y}\label{eq:dir2}
\end{eqnarray}
with $V\left(\cdot\right)$ representing an objective to be minimised and $\mathcal{U}$ and $\mathcal{Y}$ representing the input and output constraint sets, respectively.

Whereas model parameters are not explicitly derived in this formulation, an equivalence was identified in~\cite{Fiedler2020} and~\cite{Dorfler2021} between this form 
and a predictive control formulation based on a multi-step prediction model derived from data. This multi-step model version is referred to in~\cite{Dorfler2021} as an indirect data-driven formulation, in contrast to direct data-driven formulations in which no model is identified such as in (\ref{eq:dir1})--(\ref{eq:dir2}).

Using the indirect formulation, the same training data structures can be used, but here they are used to derive a multi-step predictor~$P^{*}$ by the least-squares method as
\begin{equation}
    P^{*}=\arg\min_{P}\left\|P\left[\begin{matrix}U_{\alpha}\\Y_{\alpha}\\U_{\beta}\end{matrix}\right]-Y_{\beta}\right\|^{2}_{F},
\end{equation}
where~$\|\cdot\|_{F}$ denotes the Frobenius norm. Using the Moore-Penrose inverse (denoted~$\dagger$), this can be expressed explicitly as 
\begin{equation}\label{eq:Psize}
    P^{*}:=Y_{\beta}\left[\begin{matrix}U_{\alpha}\\Y_{\alpha}\\U_{\beta}\end{matrix}\right]^{\dagger}.
\end{equation}

This predictor can then be used to derive future trajectories as
\begin{equation}\label{eq:indir}
    y_{f}=P^{*}\left[\begin{matrix}u_{ini}\\y_{ini}\\u_{f}\end{matrix}\right].
\end{equation}
With such a representation, a controller can be formulated using \eqref{eq:indir} to describe the system dynamics. Such a controller can be viewed as a form of SPC. To examine the model defined by (\ref{eq:indir}) in Section~\ref{sec:fold}, it is useful here to define a partitioned version of~$P^{*}$, given as $\left[\begin{matrix}P_{1}^{*}&P_{2}^{*}&P_{3}^{*}\end{matrix}\right]$ where $P_{1}^{*}\in\mathbb{R}^{pN\times mT_{ini}}$ is associated with the initialisation input trajectory,~$P_{2}^{*}\in\mathbb{R}^{pN\times mT_{ini}}$ is associated with the initialisation output trajectory and~$P_{3}^{*}\in\mathbb{R}^{pN\times pN}$ is associated with the future input trajectory.

An indirect data-driven predictive control formulation can then be represented by replacing (\ref{eq:dirg0}) with (\ref{eq:indir}). It should be noted that in (\ref{eq:indir}), acausal dependencies between $y_{f}$ and $u_{f}$ are permitted. To ensure causality, a lower-block triangular structure would need to be enforced on~$P_{3}^{*}$ \cite{Qin2005}. 

To aid in the design of a direct data-driven controller and to understand its performance, the authors of \cite{Dorfler2021} align the direct data-driven approach to the SPC approach via suitable regularisation and relaxation choices. In particular, the regularisation term $\left|\left|\left(I-\Pi\right)g\right|\right|^2_2$ is included in the proposed controller's objective function, where $$\Pi=\left[\begin{matrix}U_{\alpha}\\Y_{\alpha}\\U_{\beta}\end{matrix}\right]^\dagger\left[\begin{matrix}U_{\alpha}\\Y_{\alpha}\\U_{\beta}\end{matrix}\right].$$ For a noise-free, causal LTI system, trajectories that would necessitate an acausal relationship between inputs and outputs would not satisfy (\ref{eq:dirg0}), as per the fundamental lemma. However, in the presence of noise and nonlinearity, and a relaxed form of (\ref{eq:dirg0}), no such guarantee applies. Unlike the SPC case, restricting the set of allowable trajectories to those that satisfy causality conditions is not straightforward, since the link between inputs and outputs is not explicitly described. The objective of segmentation is to restructure \eqref{eq:dirg} in a manner that discourages acausal relationships without significantly increasing complexity.

In the following section, both the direct and indirect methods summarised here will be used to illustrate the rationale of a modified version of the data-predictive control approach, which is the main contribution of this work.

\subsection{Segmentation of prediction trajectory}\label{sec:fold}
In the context of the data-driven controller described in Section \ref{sec:IIA}, relaxations of the initialisation constraints and regularisation of the optimisation variables can be introduced to improve performance in this regard in the direct formulation. Similarly, slack variables can be introduced to the indirect form.

Nonetheless, the number of parameters of $P_{2}^{*}\in\mathbb{R}^{pN\times mN}$ in the indirect form increases with the horizon length. For example, the final entry of the predicted output sequence,~$y_{f}[N]$, is a function of~$N$ preceding inputs $(u_{f}[1],\ldots,u_{f}[N])$. Thus, for longer prediction horizons, the impact of over-parameterisation may become more pronounced, as will be shown in the illustrative example in Section~\ref{sec:example1}. In the direct formulation, model parameters are not explicitly derived; however, in~\cite{Fiedler2020} it is reasoned that the same model is implicitly identified in the direct form as the indirect form 
leading to the same performance drop for longer horizons. It should be noted that with perfect data, the fixing of~$u_{ini}$ and~$y_{ini}$, ensures a unique representation of~$u_{f}$ and~$y_{f}$ and thus the issue does not arise.

A modification is proposed here whereby the prediction trajectory is divided into segments of length~$T_{ini}$ (with~$T_{ini}\leq N$) to decouple the relationship between horizon length and the number of parameters (implicit parameters in the direct formulation, explicit parameters in the indirect formulation), thereby ensuring better scalability to problems with longer prediction horizons. The key insight here is that by assuming the system does not change over the prediction horizon, we can construct the full horizon using shorter trajectories. Each \textit{prediction} trajectory acts as the \textit{initialisation} trajectory for its subsequent segment.

The shorter trajectories used in this formulation necessitate a change in the training data matrix definitions. In Section~\ref{sec:II}, $T_{0}$ dictates the training length, where the conditions $T_{0}\geq(m+1)(T_{ini}+N+n)-1$ and $T_{ini}\geq \ell$ were imposed. For the segmented form, we replace~$T_{0}$ with~$T_{a}$. Since each segment is at most of length~$T_{ini}$, rather than~$N$ (the final segment may be shorter), we now impose $T_{a}\geq (m+1)(2T_{ini}+n)-1$. Notably, for~$T_{ini}<N$ a shorter training period is then sufficient. 

The updated training sequences are then defined as $u_{tr_{s}}=\left[u^{T}_{1},\ldots,u^{T}_{T_{a}}\right]^{T}\in\mathbb{R}^{mT_{a}}$ and $y_{tr_{s}}=\left[y^{T}_{1},\ldots,y^{T}_{T_{a}}\right]^{T}\in\mathbb{R}^{pT_{a}}$ and the associated Hankel matrices are defined as
\begin{equation}
    \begin{aligned}
        \left[\begin{matrix}U_{\alpha_{s}}\\U_{\beta_{s}}\end{matrix}\right]:=\mathscr{H}_{2T_{ini}}(u_{tr_{s}}),\\
        \left[\begin{matrix}Y_{\alpha_{s}}\\Y_{\beta_{s}}\end{matrix}\right]:=\mathscr{H}_{2T_{ini}}(y_{tr_{s}}),
    \end{aligned}
\end{equation}
with $U_{\alpha_{s}}\in\mathbb{R}^{mT_{ini}\times (T_{a}-2T_{ini}+1)}$, $Y_{\alpha_{s}}\in\mathbb{R}^{pT_{ini}\times (T_{a}-2T_{ini}+1)}$.

The trajectories $u_f$ and $y_f$ are partitioned into~$F$ segments given as $\left[u_{f_{1}}^{T},\ldots,u_{f_{F}}^{T}\right]^{T}=u_{f}$ and $\left[y_{f_{1}}^{T},\ldots,y_{f_{F}}^{T}\right]^{T}=y_{f}$ respectively, where $u_{f_{i}}\in\mathbb{R}^{mT_{ini}}$ and $y_{f_{i}}\in\mathbb{R}^{pT_{ini}}$, $\forall i\in\{1,\ldots,F-1\}$, and the final segments $u_{f_{F}}\in\mathbb{R}^{m(N-(F-1)T_{ini})}$ and $y_{f_{F}}\in\mathbb{R}^{p(N-(F-1)T_{ini})}$. For notational brevity,~$u_{ini}$ and~$y_{ini}$ are replaced by~$u_{f_{0}}$ and~$y_{f_{0}}$, respectively.

\begin{figure}[!tb]
\centering
\begin{subfigure}[b]{\mysize\textwidth}
    \centering
    \includegraphics[width=\textwidth]{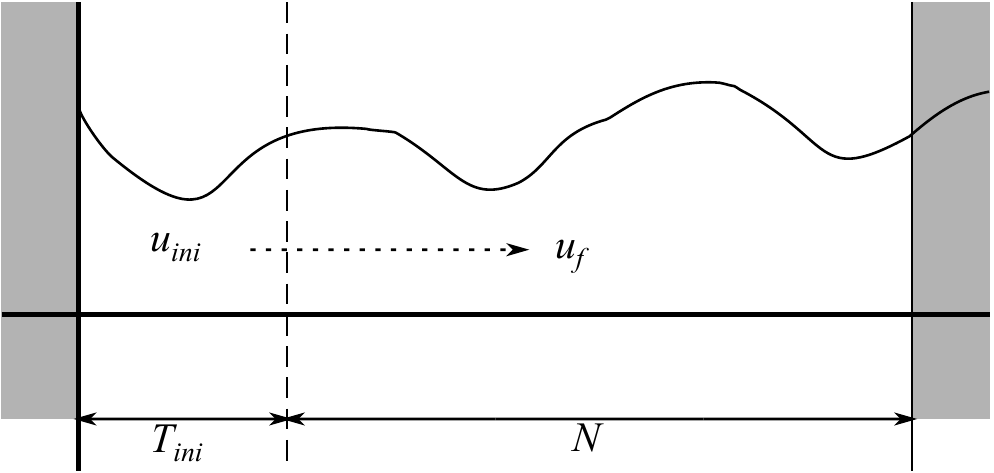}
    \caption{}
    \label{fig:Line0}
\end{subfigure}
\hfill
\begin{subfigure}[b]{\mysize\textwidth}
    \centering
    \includegraphics[width=\textwidth]{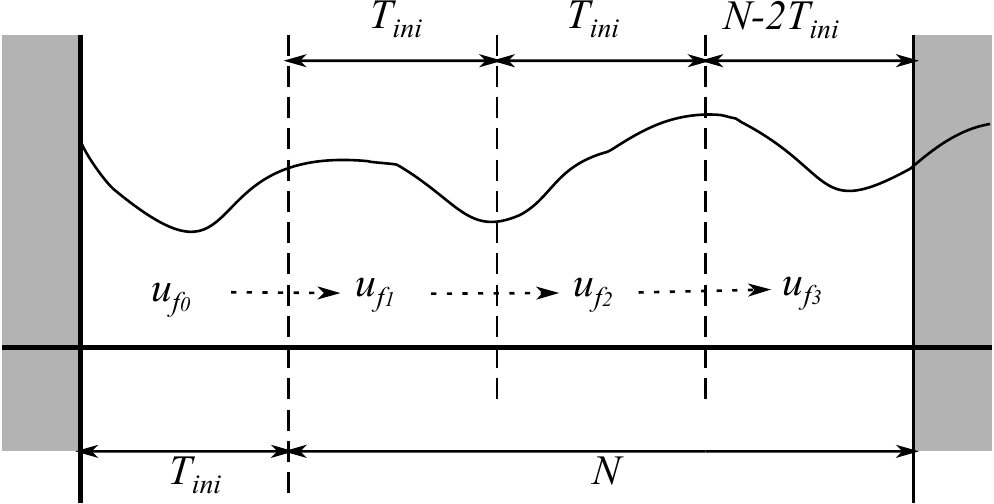}
    \caption{}
    \label{fig:Line1}
\end{subfigure}
\caption{Illustration of data-driven control approach (a) without and (b) with segmentation of prediction trajectory}
\label{fig_linediag}
\end{figure}

A diagram illustrating the segmentation concept is shown in Fig.~\ref{fig_linediag} for a prediction trajectory divided into three segments.

The indirect approach can now be reformulated. The new multi-step predictor matrix, denoted ~$P_{s}^{*}$, can be found as
\begin{equation}
    P_{s}^{*}=Y_{\beta_{s}}\left[\begin{matrix}U_{\alpha_{s}}\\Y_{\alpha_{s}}\\U_{\beta_{s}}\end{matrix}\right]^{\dagger}.
\end{equation}
As in the unsegmented case, $P_{s}^{*}$ can be partitioned and each of the $F$ segments of the prediction trajectory can then be represented as:
\begin{equation}\label{eq:indFirst}
    y_{f_{i}}=\left[P_{s_1}^{*},P_{s_2}^{*},P_{s_3}^{*}\right]\left[\begin{matrix}u_{f_{i-1}}\\y_{f_{i-1}}\\u_{f_{i}}\end{matrix}\right],\forall i\in \{1\ldots,F\}.
\end{equation}
If the length~$N$ of the desired trajectory is not a multiple of~$T_{ini}$, i.e. if~$T_{ini}F$ is longer than~$N$, the final~$T_{ini}F-N$ terms of the final segment~$F$ can be ignored.

To observe the structure of the predictor over the full horizon in the same form as (\ref{eq:indir}), the segments can be stacked and rearranged, resulting in:
\begin{eqnarray}\label{eq:indiestack}
    \left[\begin{matrix}y_{f_1}\\y_{f_2}\\\vdots\\y_{f_F}\end{matrix}\right]=I_{F}\otimes P^{*}_{s_1}\left[\begin{matrix}u_{f_0}\\u_{f_1}\\\vdots\\u_{f_{F-1}}\end{matrix}\right]+I_{F}\otimes P^{*}_{s_3}\left[\begin{matrix}u_{f_1}\\u_{f_2}\\\vdots\\u_{f_F}\end{matrix}\right]\nonumber\\
    +I_{F}\otimes P^{*}_{s_2}\left[\begin{matrix}y_{f_0}\\y_{f_1}\\\vdots\\y_{f_{F-1}}\end{matrix}\right].
\end{eqnarray}

This can be represented more concisely as:
\begin{equation}
y_{f}=\left[\begin{matrix}\Phi_{1}&\Phi_{2}&\Phi_{3}\end{matrix}\right]\left[\begin{matrix}u_{ini}\\y_{ini}\\u_{f}\end{matrix}\right],
\end{equation}
where $\Phi_{1}$, $\Phi_{2}$ and $\Phi_{3}$ are the segmented counterparts of $P_{1}^{*}$, $P_{2}^{*}$ and $P_{3}^{*}$, with the same dimensions. In this form, $\Phi_{3}$ is lower-block triangular, given as:
\begin{equation}
    \left[\begin{matrix}P^{*}_{s_3}&0&\cdots&0\\P^{*}_{s_3}\!P^{*}_{s_2}\!+\!P^{*}_{s_1}&P^{*}_{s_3}&\ddots&\vdots\\\vdots&\ddots&\ddots&0\\P^{*}_{s_3}\!P^{*\!^{F\!-\!1}}_{s_2}\!+\!P^{*}_{s_1}\!P^{*\!^{F\!-\!2}}_{s_2}&\cdots&P^{*}_{s_3}\!P^{*}_{s_2}\!+\!P^{*}_{s_1}&P^{*}_{s_3}\end{matrix}\right].
\end{equation}

Notably, $\Phi_{3}$ is lower-block triangular, implying that any output segment cannot be influenced by inputs from future segments. Although this does not ensure causality within segments, since outputs can be influenced by future inputs from the same segment, these can be at most $T_{ini}$ samples ahead. In contrast, without segmentation outputs can be influenced by inputs from the full horizon. The impact of the restructuring on $\Phi_{3}$ is shown in the numerical example of Section~\ref{sec:example1}.

Next, the segmented structure is translated to the direct formulation in the same manner as Section~\ref{sec:IIA}. Equation~\ref{eq:dirg0} can be replaced by the following:

\begin{equation}
    \left[\begin{matrix}U_{\alpha_{s}}\\U_{\beta_{s}}\\Y_{\alpha_{s}}\\Y_{\beta_{s}}\end{matrix}\right]g_{i}=\left[\begin{matrix}u_{f_{i-1}}\\u_{f_{i}}\\y_{f_{i-1}}\\y_{f_{i}}\end{matrix}\right],\quad\forall i\in\{1,\ldots,F\}\label{eq:segConst0},
\end{equation}
where $g_{i}\in\mathbb{R}^{T_{a}-2T_{ini}+1}$, $\forall i\in\{1,\ldots,F\}$.

Using (\ref{eq:segConst0}) to predict future input and output trajectories of the system, we can formulate a predictive controller. Here, we seek to minimise a cost function given as~$V_{s}(\cdot)$ along with the regularisation term given as $\left|\left|\left(I-I_{F}\otimes\Pi_{s}\right)g\right|\right|^2_2$ where~$\otimes$ denotes the Kronecker product and~$I_{j}$ denotes the identity matrix of~size $j\times j$. The objective is then
\begin{equation}
    \min_{g_{1},\ldots,g_{F}}\sum_{i=1}^{F}V_{s}(g_{i})+\left|\left|\left(I-I_{F}\otimes\Pi_{s}\right)g\right|\right|^2_2.
\end{equation}
The constraints are given as:
\begin{eqnarray}
    \left[\begin{matrix}U_{\alpha_{s}}\\Y_{\alpha_{s}}
    \end{matrix}\right]g_{1}&=&\left[\begin{matrix}u_{f_{0}}\\y_{f_{0}}\end{matrix}\right]\\
    \left[\begin{matrix}-U_{\beta_{s}}&U_{\alpha_{s}}\\-Y_{\beta_{s}}&Y_{\alpha_{s}}\end{matrix}\right]\left[\begin{matrix}g_{i-1}\\g_{i}\end{matrix}\right]&=&\boldsymbol{0}_{2T_{ini}},\forall i\in\{2,\ldots,F\}\\
    U_{\beta_{s}}g_{i}&\in&\mathcal{U},\forall i\in\{1,\ldots,F\}\\
    Y_{\beta_{s}}g_{i}&\in&\mathcal{Y},\forall i\in\{1,\ldots,F\}
\end{eqnarray}
where $\boldsymbol{0}_{a}$ denotes a column of zeros of length~$a$. No penalty on the input is included. Since the length of $g$ in the unsegmented form is at least $m\left(T_{ini}+N+n\right)+n$ and the length of $\left[g_1,\ldots,g_F\right]$ for the segmented formulation is at least $F\left(m\left(2T_{ini}+n\right)+n\right)$, the number of decision variables increases with segmentation. Nonetheless, the block-diagonal structure of the Hessian matrix in the segmented problem implies that benefits can be achieved through the use of sparse quadratic programming solvers, leading to a linear increase in computational time with increasing horizon length. In Section \ref{sec:example1}, this is shown empirically for the segmented and unsegmented forms.

\section{Illustrative example: Two-mass system}\label{sec:example1}

\subsection{System description}
\begin{figure}[!tb]
\centering
\includegraphics[width=\mysize\textwidth]{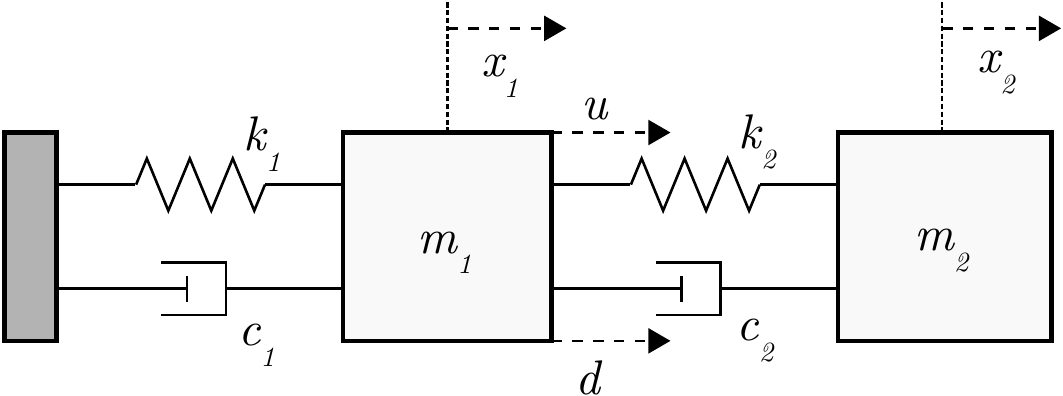}
\caption{Two-mass system with springs and dampers}
\label{fig_2mass}
\end{figure}

A two-mass-spring-damper example is used to illustrate the performance of the segmented predictive controller compared with the unsegmented version. The code needed to reproduce these examples is available on Code Ocean. The system comprises two masses, two springs and two dampers, and is described in the following equations:
\begin{eqnarray}
    \dot{x}(t)=&{}Ax(t)+B\left(u(t)+d(t)\right)\\
    y(t)=&{}Cx(t)+\upsilon(t),
\end{eqnarray}
where $x=(x_{1},x_{2},\dot{x}_{1},\dot{x}_{2})$ with~$x_{1}$ and~$x_{2}$ representing the displacement of masses~$m_{1}$ and~$m_{2}$ respectively, and their corresponding velocities (shown in Fig~\ref{fig_2mass}). Then~$u(t)$ is the input force applied to the mass~$m_{1}$ and $y(t)$ is the measured displacement of $m_{2}$. An additional disturbance~$d(t)$ can be applied to~$m_{1}$ as well as measurement noise~$\upsilon(t)$. The observation matrix~$C:=\left[\begin{matrix}0&1&0&0\end{matrix}\right]$, while the parameter matrices~$A$ and~$B$ are given as
\begin{equation}
    A:=\left[\begin{matrix}0 & 0 & 1 & 0\\
    0 & 0 & 0 & 1\\
    \frac{-\left(k_{1}+k_{2}\right)}{m_{1}} & \frac{k_{2}}{m_{1}} & \frac{-\left(c_{1}+c_{2}\right)}{m_{1}} & \frac{c_{2}}{m_{1}}\\
    \frac{k_{2}}{m_{2}} & \frac{-k_{2}}{m_{2}} & \frac{c_{2}}{m_{2}} & \frac{-c_{2}}{m_{2}}\end{matrix}\right],B:=\left[\begin{matrix}0\\0\\\frac{1}{m_{1}}\\0\end{matrix}\right],
\end{equation}
where the masses are defined~$m_{1}=0.5$ and~$m_{2}=1.5$, the spring constants are defined as~$k_{1}=2$ and $k_{2}=2$ and the damping constants are defined as~$c_{1}=1$ and~$c_{2}=1$. The system is shown in Fig.~\ref{fig_2mass}.

\subsection{Problem formulation}
\label{sec:MassesProblem}
We investigate a case study using this system whereby we seek to control the displacement~$y$ to track a set-point~$y_{sp}$, by calculating an input trajectory~$u$ using a data-driven predictive controller. We adopt a direct data-driven formulation and compare segmented and unsegmented versions of the strategy in scenarios with different realisations of a time-varying unmeasured disturbance~$d$ applied to~$m_1$ and different realisations of measurement noise $\upsilon$. A one-second sample time is used for the predictive controller, with input and disturbance signals held constant for the duration of the sample. The input force is constrained to the interval $\left[-1,1\right]$ and results are compiled from a 100-second run.

\subsection{Prioritised objective formulation}
To handle the regularisation, relaxation and set-point deviation penalties, a prioritised framework is used as described in~\cite{Kerrigan2002}. The tracking problem from Section \ref{sec:MassesProblem} is solved in two stages whereby a feasibility stage is followed by a set-point deviation minimisation stage. The first optimisation minimises the initialisation slacks given as $\varepsilon_{f_{i}}\in\mathbb{R}^{T_{ini}}$, $\forall i\in \{1,\ldots,F\}$. The objective and constraints of this linear problem are defined as follows:

\begin{equation}\label{eq:obj1}
    J^{*}_{1}:=\min_{\scriptsize{\begin{array}{c}g_{1}\ldots,g_{F},\\\varepsilon_{f_{1}},\ldots,\varepsilon_{f_{F}}\end{array}}}\sum_{i=1}^{F}\sum_{j=1}^{T_{ini}}\varepsilon_{f_{i,j}}
\end{equation}
$\qquad\qquad \text{s.t.}$
\begin{eqnarray}
    U_{\alpha_{s}}g_{1}
    =&u_{f_{0}}\label{eq:c1}\\
    \left[\begin{matrix}Y_{\alpha_{s}}\\-Y_{\alpha_{s}}\end{matrix}\right]g_{1}-\left[\begin{matrix}\varepsilon_{f_{1}}\\\varepsilon_{f_{1}}\end{matrix}\right]
    \leq&\left[\begin{matrix}y_{f_{0}}\\-y_{f_{0}}\end{matrix}\right]\label{eq:c2}\\
    \left[\begin{matrix}-U_{\beta_{s}}&U_{\alpha_{s}}\end{matrix}\right]\left[\begin{matrix}g_{i-1}\\g_{i}\end{matrix}\right]
    =&\boldsymbol{0}_{T_{ini}},\forall i\in\{2,\ldots,F\}\label{eq:c3}\\
    \left[\begin{matrix}-Y_{\beta_{s}}&Y_{\alpha_{s}}\end{matrix}\right]\left[\begin{matrix}g_{i-1}\\g_{i}\end{matrix}\right]-
    \varepsilon_{f_{i}}
    \leq&\boldsymbol{0}_{T_{ini}},\forall i\in\{2,\ldots,F\}\label{eq:c4}\\
    \left[\begin{matrix}Y_{\beta_{s}}&-Y_{\alpha_{s}}\end{matrix}\right]\left[\begin{matrix}g_{i-1}\\g_{i}\end{matrix}\right]-
    \varepsilon_{f_{i}}
    \leq&\boldsymbol{0}_{T_{ini}},\forall i\in\{2,\ldots,F\}\label{eq:c5}\\
    -\varepsilon_{f_{i}}
    \leq&\boldsymbol{0}_{T_{ini}},\forall i\in\{1,\ldots,F\}\label{eq:c6}\\
    U_{\beta_{s}}g_{i}
    \in&\mathcal{U},\forall i\in\{1,\ldots,F\}\\
    Y_{\beta_{s}}g_{i}
    \in&\mathcal{Y},\forall i\in\{1,\ldots,F\}\label{eq:c7}.
\end{eqnarray}

The second optimisation objective is composed of a penalty on the sum of the absolute deviation of the output from the set-point, given as $\varepsilon_{y}\in\mathbb{R}^{N}$, and a regularisation penalty on~$g$ with the relative weight between the two penalties set by choice of~$\lambda_{g}>0$. The quadratic objective and linear constraints of this problem are given as
\begin{equation}
    J_{2}^{*}:=\min_{\scriptsize{\begin{array}{c}g_{1}\ldots,g_{F},\varepsilon_{y},\\\varepsilon_{f_{1}},\ldots,\varepsilon_{f_{F}}\end{array}}}\sum_{j=1}^{N}\varepsilon_{y_{j}}+\lambda_{g}\left|\left|\left(I-I_{F}\otimes\Pi_{s}\right)g\right|\right|^2_2
\end{equation}
$\qquad\qquad \text{s.t.}$
\begin{eqnarray}
    (\ref{eq:c1})\varA{--}(\ref{eq:c7})\nonumber\\
    \sum_{i=1}^{F}\sum_{j=1}^{T_{ini}}\varepsilon_{f_{i,j}}&\leq& J^{*}_{1}\\
    \left[\begin{matrix}I_{F}\otimes Y_{\beta_{s}}\\-I_{F}\otimes Y_{\beta_{s}}\end{matrix}\right]\left[\begin{matrix}g_{1}\\\vdots\\g_{F}\end{matrix}\right]-\left[\begin{matrix}\varepsilon_{y}\\\varepsilon_{y}\end{matrix}\right]&\leq& \left[\begin{matrix}y_{sp}\\-y_{sp}\end{matrix}\right]\\
    -\varepsilon_{y}&\leq& \boldsymbol{0}_{N}.\label{eq:objend}
\end{eqnarray}

For the unsegmented formulation, the same formulation is used with $F=1$, constraints (\ref{eq:c3}), (\ref{eq:c4}) and (\ref{eq:c5}) omitted and $\alpha_{s}$ and $\beta_{s}$ subscripts replaced by $\alpha$ and $\beta$ respectively.

To generate training data, the system was simulated in open-loop with the input force $u$ varied at 10-second intervals by drawing a sample from a uniform distribution in the interval $[-1$ N$,1$ N$]$ to generate a persistently exciting training set of input and output data. A disturbance signal, which acts as the unmeasured disturbance, was also generated and applied to the system. This disturbance force was composed of a sinusoidal component of amplitude 0.2 N, a bias of 0.2 N and a frequency of 0.01 Hz added to a uniformly distributed random noise component drawn from the interval~$[-0.15$ N$,0.15$ N$]$. The measurement noise was drawn from a normal distribution with zero mean and a standard deviation of 0.1 m.

\subsection{Performance analysis: disturbance and measurement noise}
To illustrate the behavioural differences between the segmented and unsegmented formulations, the system is simulated for 100 different realisations of the stochastic disturbance term followed by 100 different realisations of the measurement noise term for several prediction horizon lengths. The two formulations are compared by observing the set-point deviations achieved for each formulation across the different horizon choices for each noise and disturbance realisation. The scenarios with unmeasured disturbance present are first shown, followed by the scenarios with measurement noise present. The benefits of the segmented approach are then discussed. 

\subsubsection{Tuning}
In all cases, the regularisation weight $\lambda_{g}=0.5$ and initialisation length $T_{ini}=5$. It should be noted that, though the outcomes were not sensitive to these choices in this case study, parameter selection is not necessarily a trivial task, as indicated in \cite{Data_Chinde2022}. In this work, the choices were made by evaluating the set-point tracking performance of both segmented and unsegmented formulation for different parameter values by trial-and-error. A more rigorous tuning approach that does not require a model would be needed for real-world implementation purposes. This is considered outside the scope of this work, but automated tuning methods have been developed with this in mind \cite{ODwyerEcos}.

\subsubsection{Disturbance realisations}

\begin{figure}[!tb]
    \centering
    \includegraphics[width=\mysize\textwidth]{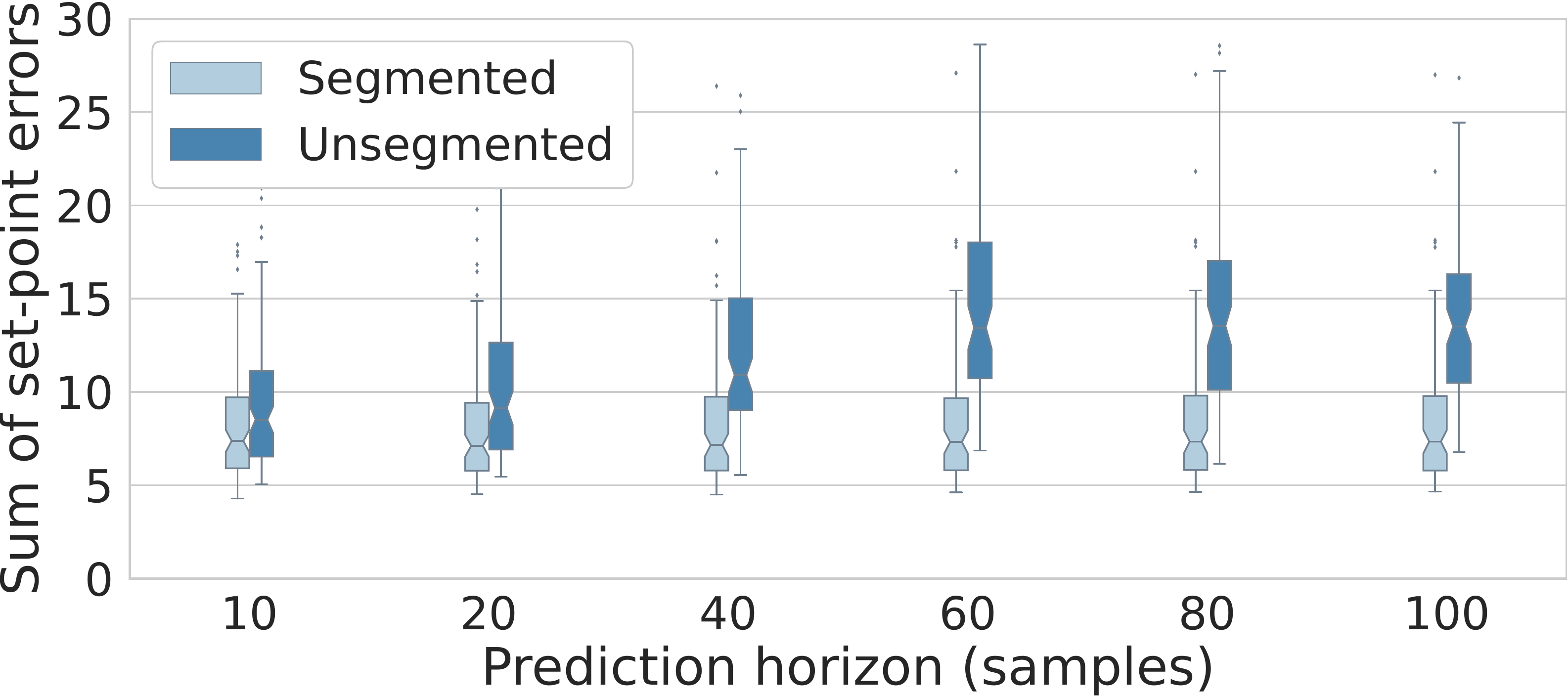}
    \caption{Performance of segmented and unsegmented formulations for different realisations of the unmeasured disturbance across different prediction horizons}
    \label{fig:boxplot}
\end{figure}

In Fig.~\ref{fig:boxplot}, the results for each realisation of the disturbance are shown in the form of a box plot in which results are grouped in terms of horizon length. In the plot, a notch is centred on the median and the lower and upper sides of the boxes themselves show the lower and upper quartiles respectively.

\begin{figure}[!tb]
\centering
\includegraphics[width=\mysize\textwidth]{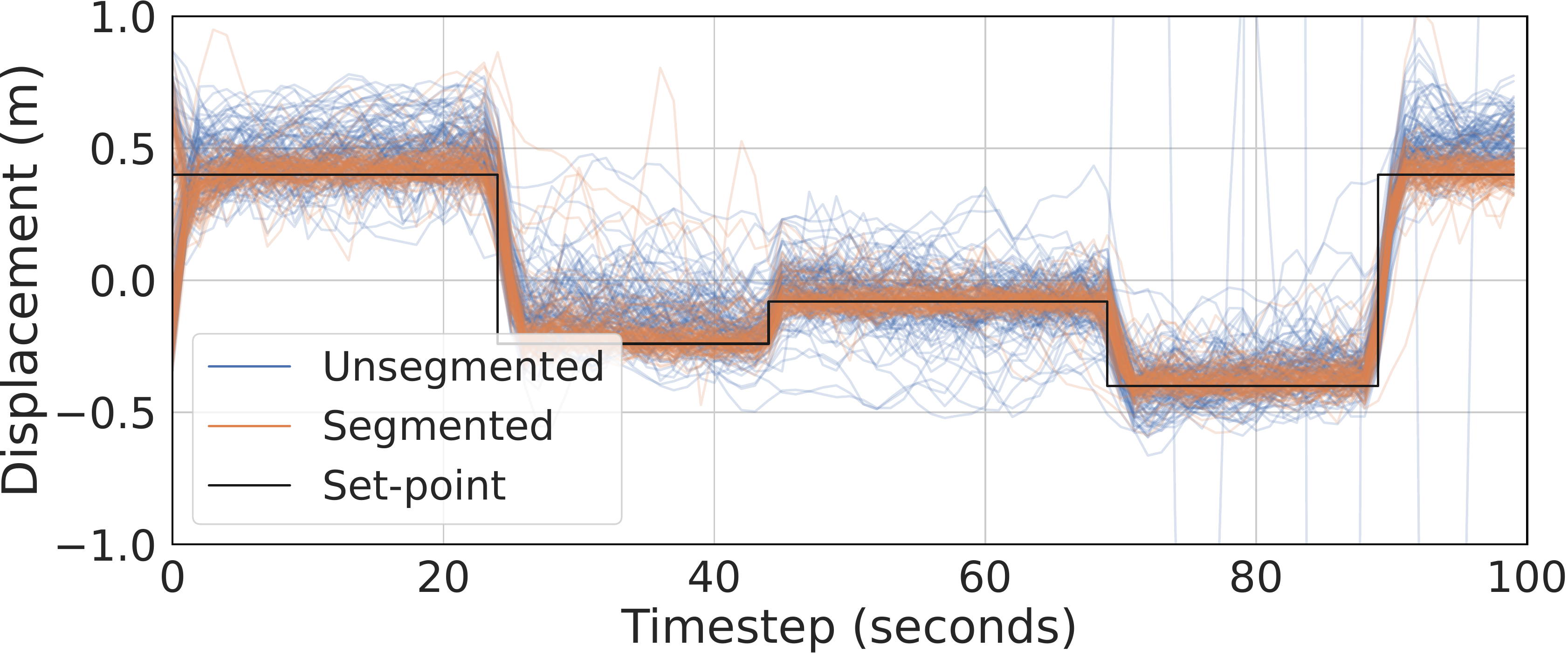}
\caption{Output of 2-mass system for each disturbance realisation with $N=100$ for segmented and unsegmented formulations}
\label{fig_linesplot0}
\end{figure}

It can be seen that the segmented formulation tends to outperform the unsegmented version, with the performance more pronounced in longer horizons. The segmented formulation reduced the impact of the disturbance on the optimisation problem, as will be further presented in Section \ref{sec:Benefits}, enabling better tracking accuracy. Notably, the performance of the segmented formulation is unaffected by the choice of prediction horizon. The difference in performance is illustrated in Fig~\ref{fig_linesplot0} in which the output of the system is plotted for each realisation with $N=100$ for both segmented and unsegmented. The segmented outputs are generally grouped more closely to the set-point than those of the unsegmented formulation.

By comparing the outcomes of each disturbance realisation individually, the results are summarised in Table~\ref{table_box1}. The table shows two metrics. The first, labelled \textit{Outperforming scenarios}, is the percentage of realisations for which segmentation led to a lower set-point error, with set-point error defined as the sum of the distances of the output to the set-point across the full simulation. The second, \textit{Average performance improvement}, shows the average percentage set-point error reduction achieved through segmentation across all disturbance realisations. For longer horizons, segmentation improved performance in approximately 85\% of cases, with an average improvement of over 30\%.

\begin{table}[!tb]
\renewcommand{\arraystretch}{1.3}
\caption{Improvement of segmented versus unsegmented for external disturbance scenarios}
\label{table_box1}
\centering
    \begin{tabular}{c|c|c}
     & Outperforming & Average performance\\
     & scenarios & improvement\\
      & (\% scenarios) & (\% error reduction)\\
    \hline
    \hline
    $N=10$ & 69\% & 13\% \\
    \hline
    $N=20$ & 74\% & 17\% \\
    \hline
    $N=40$ & 83\% & 28\% \\
    \hline
    $N=60$ & 88\% & 36\% \\
    \hline
    $N=80$ & 85\% & 34\% \\
    \hline
    $N=100$ & 86\% & 32\% \\
    \hline
    \end{tabular}
\end{table}

\subsubsection{Measurement noise realisations}

\begin{figure}[!tb]
    \centering
    \includegraphics[width=\mysize\textwidth]{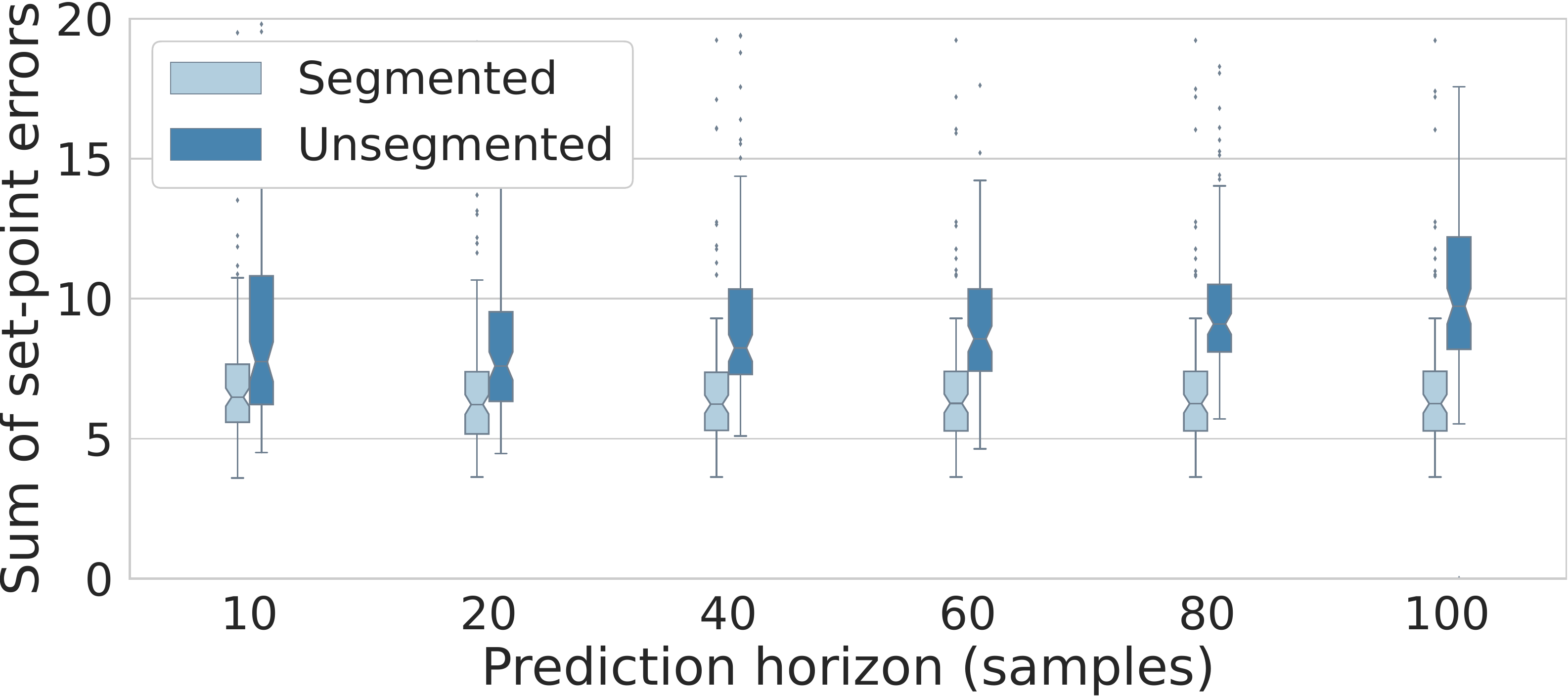}
    \caption{Performance of segmented and unsegmented formulations for different realisations of the measurement noise across different prediction horizons}
    \label{fig:boxplotsens}
\end{figure}

Next, the results for different realisations of the measurement noise are shown in Fig.~\ref{fig:boxplotsens}. Once again, segmentation leads to performance benefits in terms of set-point error reduction and the segmented performance is consistent across prediction horizon lengths.

For longer horizons, segmentation led to a performance improvement in over 85\% of the simulated cases with an average reduction in set-point error of over 30\%. These results are summarised in Table~\ref{table_box2}.

\begin{table}[!tb]
\renewcommand{\arraystretch}{1.3}
\caption{Improvement of segmented versus unsegmented for sensor noise scenarios}
\label{table_box2}
\centering
    \begin{tabular}{c|c|c}
     & Outperforming & Average performance\\
     & scenarios & improvement\\
      & (\% of scenarios) & (\% error reduction)\\
    \hline
    \hline
    $N=10$ & 77\% & 30\% \\
    \hline
    $N=20$ & 76\% & 17\% \\
    \hline
    $N=40$ & 83\% & 26\% \\
    \hline
    $N=60$ & 83\% & 25\% \\
    \hline
    $N=80$ & 85\% & 28\% \\
    \hline
    $N=100$ & 86\% & 33\% \\
    \hline
    \end{tabular}
\end{table}

\subsubsection{Benefits of segmentation}
\label{sec:Benefits}
\begin{figure}[!tb]
\centering
    \begin{subfigure}[b]{0.24\textwidth}
    \includegraphics[width=\textwidth]{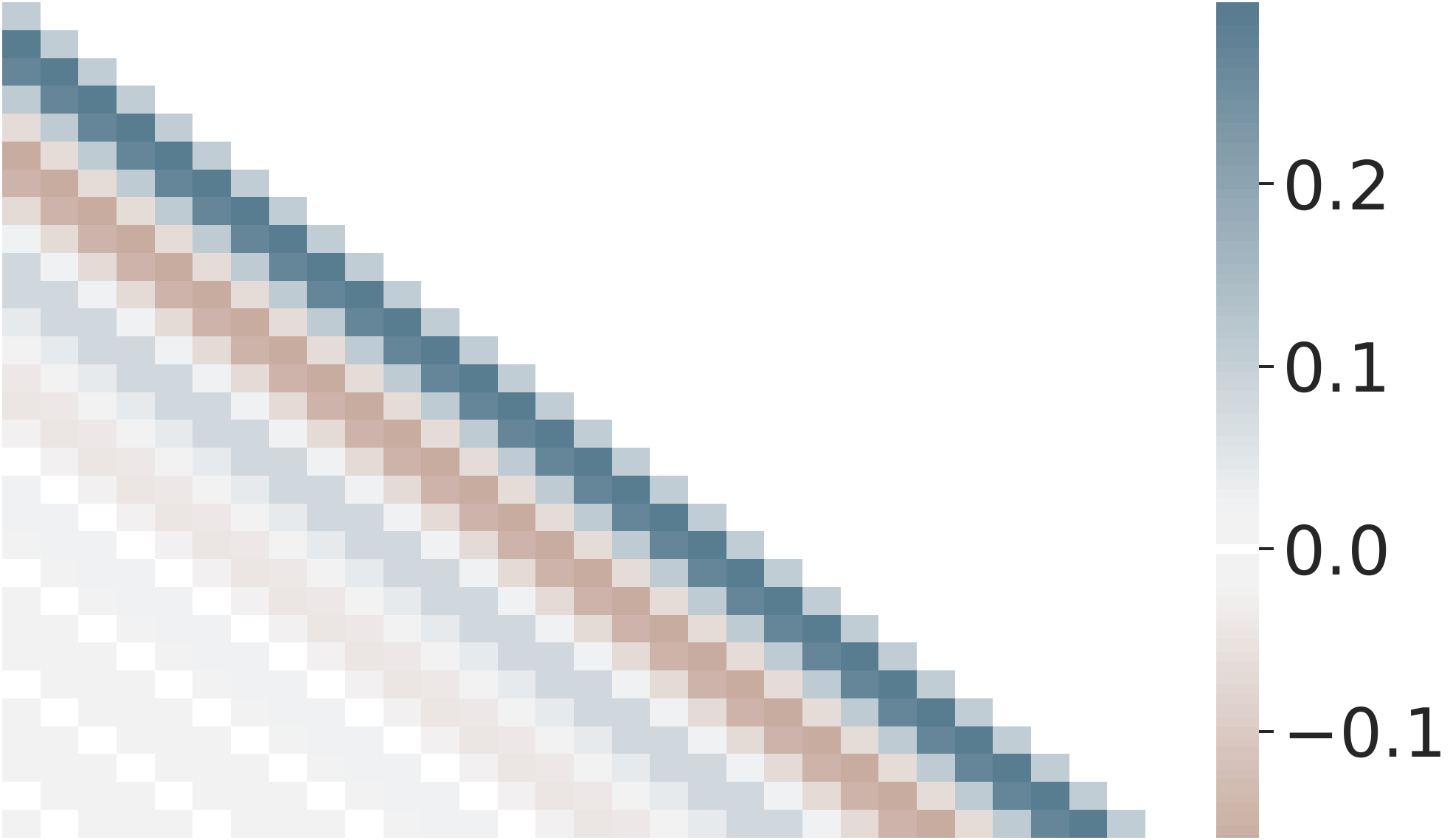}
    \caption{}
    \label{fig:HMapUnseg}
    \end{subfigure}
    \begin{subfigure}[b]{0.24\textwidth}
    \includegraphics[width=\textwidth]{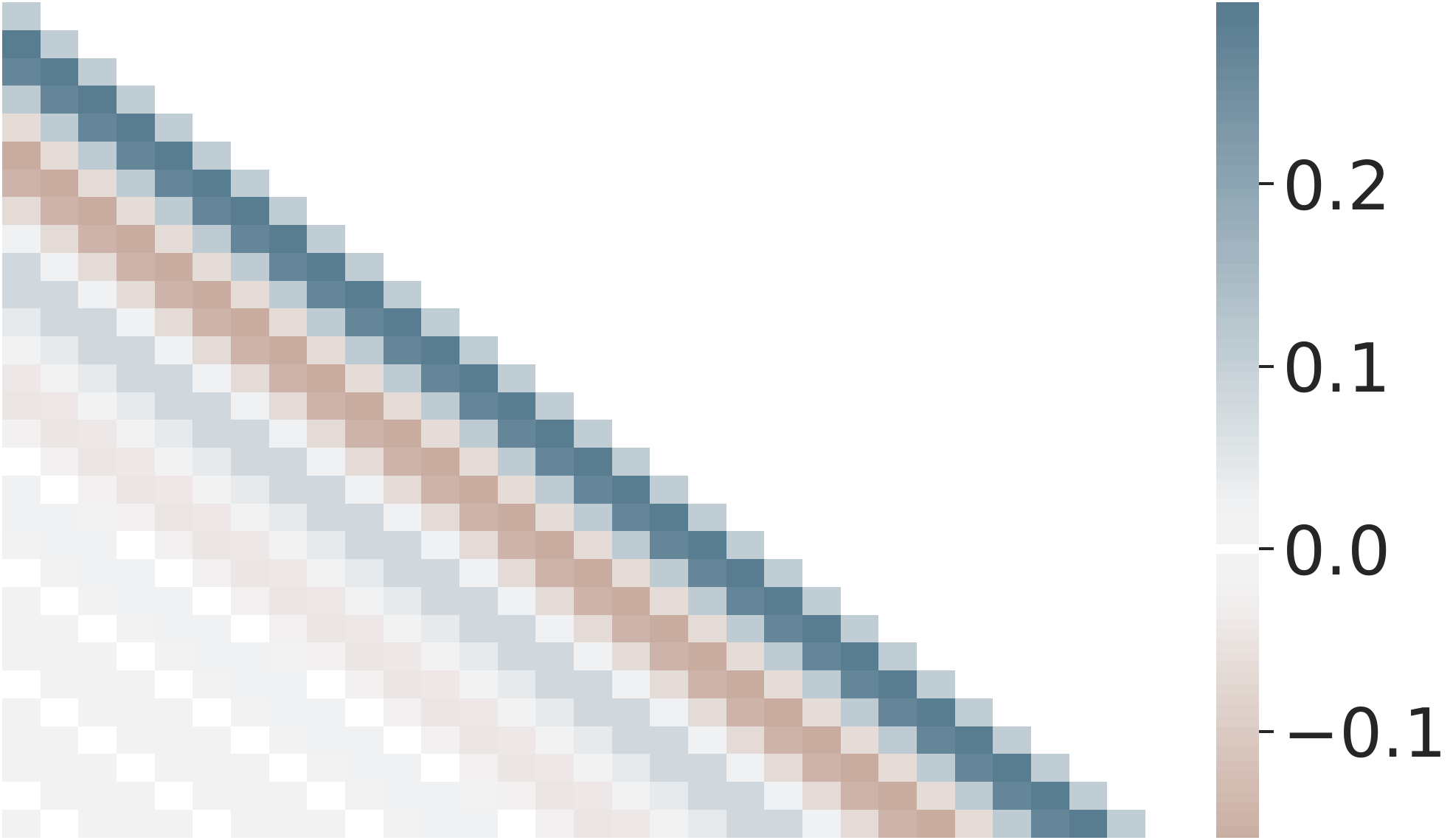}
    \caption{}
    \label{fig:HMapSeg}
    \end{subfigure}
    \begin{subfigure}[b]{0.24\textwidth}
    \includegraphics[width=\textwidth]{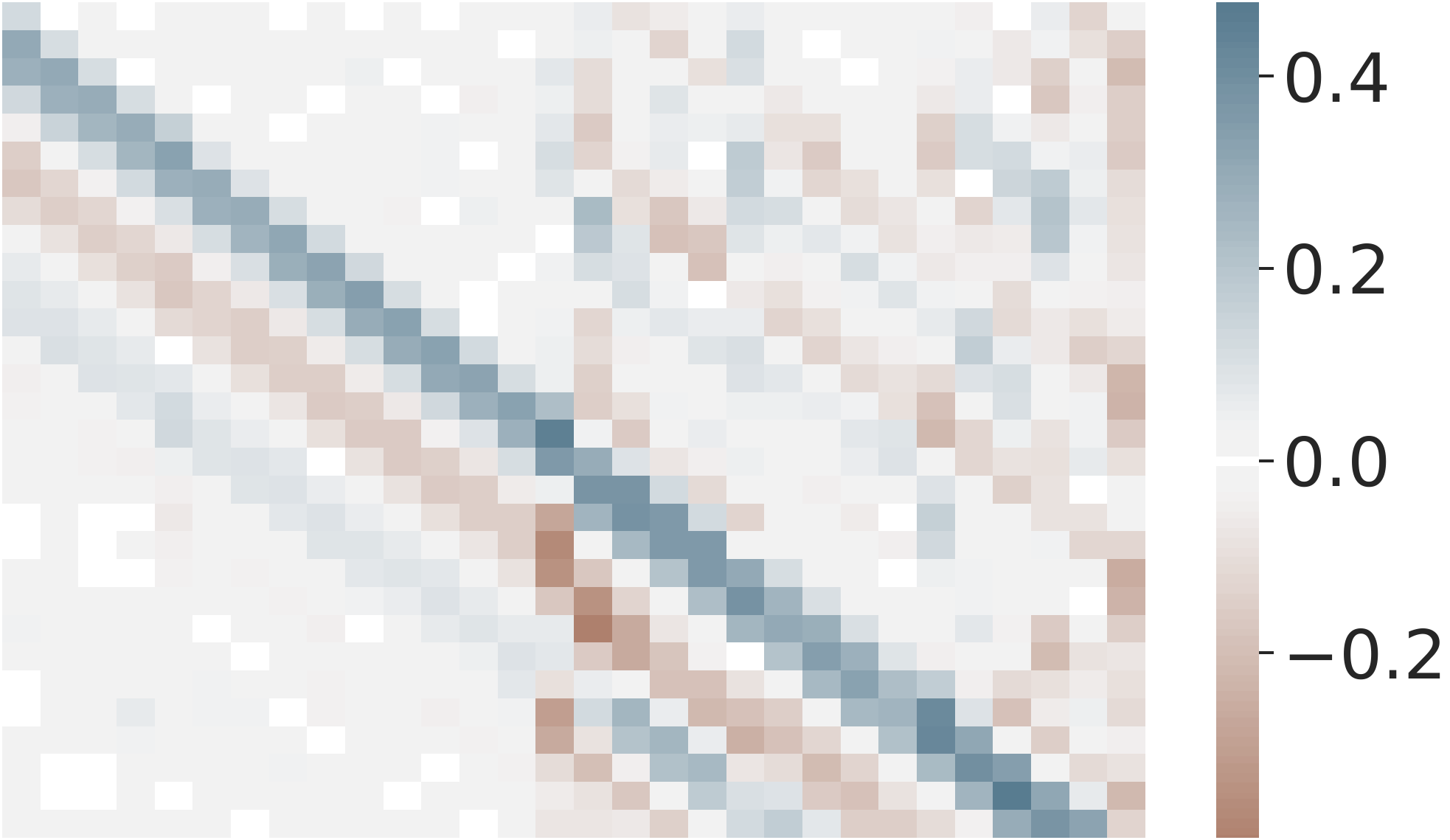}
    \caption{}
    \label{fig:HMapUnsegNoise}
    \end{subfigure}
    \begin{subfigure}[b]{0.24\textwidth}
    \includegraphics[width=\textwidth]{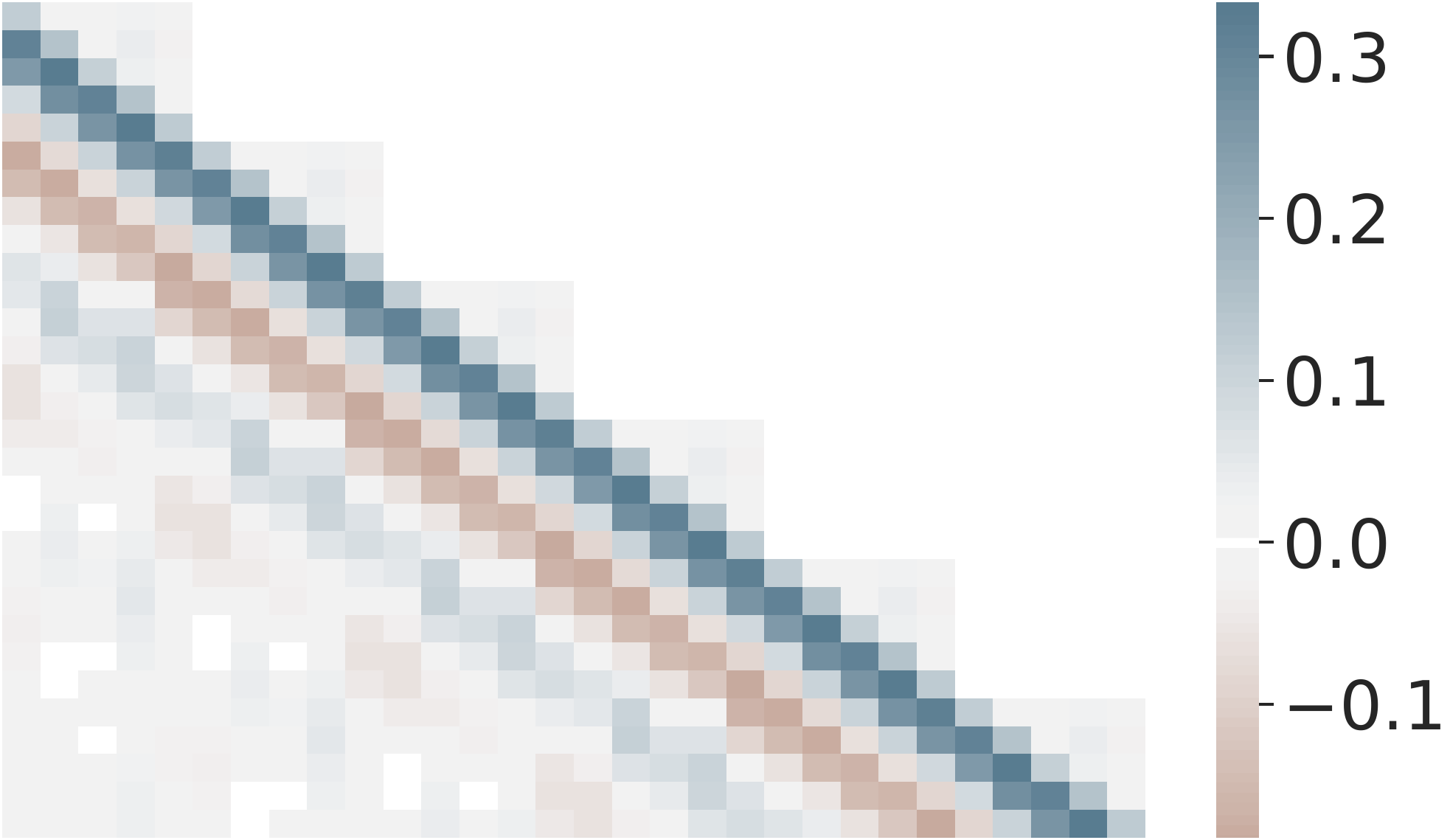}
    \caption{}
    \label{fig:HMapSegNoise}
    \end{subfigure}
\caption{Heat maps showing the composition of (a) $P^{*}_{3}$ unsegmented, no disturbance; (b) $\Phi_{3}$ segmented, no disturbance; (c) $P^{*}_{3}$ unsegmented, with disturbance; (d) $\Phi_{3}$ segmented, with disturbance. Notice that the upper-triangular portion of (c) is non-zero.}
\label{fig_Heatmaps}
\end{figure}

To understand the performance benefits of segmentation, it is instructive to consider the discussion of Section~\ref{sec:fold} in the context of the case study results. It was mentioned in Section~\ref{sec:fold} that causality entails a block-diagonal structure on $P^{*}_{3}$ and $\Phi_{3}$ for unsegmented and segmented formulations respectively. Though these matrices are related to an SPC formulation rather than the direct DeePC formulation, the latter is regularised to align with the former, making their structures relevant to DeePC. By visualising the values of these square matrices taken from the simulated examples in the form of heat maps, the impact of the disturbance can be observed on each. Such heat maps are shown in Fig.~\ref{fig_Heatmaps} for unsegmented and segmented formulations, with and without disturbance for a case with $N=30$. Without disturbance, $P^{*}_{3}$ and $\Phi_{3}$ are very similar, as shown in Fig.~\ref{fig:HMapUnseg}-\ref{fig:HMapSeg}. In the disturbed cases, $\Phi_{3}$ taken from the segmented case (Fig.~\ref{fig:HMapSegNoise}) is similar to the undisturbed cases, whereas the unsegmented cases shown in Fig.~\ref{fig:HMapUnsegNoise} is quite different. Of particular note is the presence of non-zero terms in the upper-triangular portion of the matrix.

\subsection{Performance analysis: Computation time}
\begin{figure}[!tb]
\centering
\includegraphics[width=\mysize\textwidth]{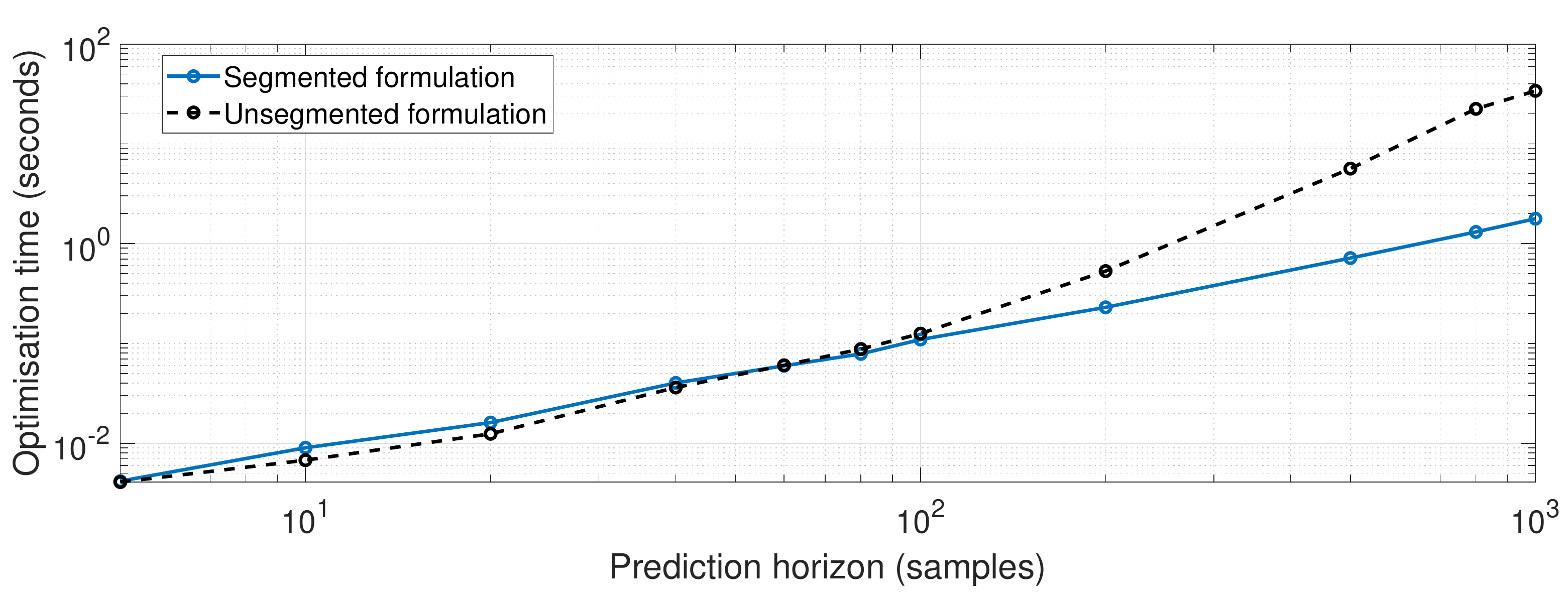}
\caption{Computational time for segmented and unsegmented formulations}
\label{fig_timeplot}
\end{figure}

As the problem structure of the segmented formulation is different to the unsegmented formulation, it is worth considering the computational time required for each. Without segmentation, the number of decision variables for the case study is $2N+30$. With segmentation, this becomes $8N$, which is larger than the unsegmented form for $N>5$. Despite the increase in decision variables, the block-diagonal structure of the Hessian in the segmented problem allows for the problem sparsity to be exploited. This was examined by observing the computation time needed to solve the second-level quadratic optimisation with various prediction horizons. Parameter choices of $\lambda_{g}=0.5$ and $T_{ini}=5$ were used for all cases. All scenarios were computed using the \textit{quadprog} function, with the interior-point-convex algorithm, using the sparse setting for the internal linear solver in Matlab 2020b on a 2.9 GHz processor, with the results shown in the log-log plot in Fig.~\ref{fig_timeplot}. For shorter horizons, the solution time of the segmented formulation is slightly higher, while the converse is true for longer horizons ($N\geq60$). Indeed the computation time of the segmented formulation increases approximately linearly with increasing horizon length, while the computation time of the unsegmented formulation is nonlinear.

\section{Application to building energy management}\label{sec_bldg}
\subsection{The building energy management challenge}
An active area of research in recent times concerns the use of predictive control for building energy management. Modern energy systems require more flexibility to handle the combined influences of increased renewable generation and increased electrification of heating and transport. Making use of buildings as active, flexible components in such an energy landscape is a key requirement in global decarbonisation efforts~\cite{Lowes2020}. Despite the pressing need for advanced control technologies, the underlying model complexity of a building and the wide variation in building designs has led to the model development process acting as a significant barrier to technology uptake~\cite{Atam2016}. Consequently, data-driven predictive control techniques have recently received attention for the application of building energy management~\cite{Kathirgamanathan2021}. 

The segmented formulation proposed in this paper is suited to this domain. Diurnal building usage and energy tariff patterns, along with the slow thermal dynamics of well-insulated buildings, make longer prediction horizons advantageous. Furthermore, many disturbances tend to impact the energy demand of a building. Measurements of these may not be available. External temperature, solar radiation and internal gains will influence the building's behaviour, potentially corrupting the ability of a data-driven algorithm to identify input/output behaviour from a given data-set. A simulated case study was carried out to investigate the performance of the segmented formulation in this setting, using state-of-the-art EnergyPlus~\cite{DOE2017} building simulation software and comparing the performance of the unsegmented and segmented formulations.

\subsection{Building simulation environment}
\begin{figure}[!tb]
\centering
\includegraphics[width=\mysize\textwidth]{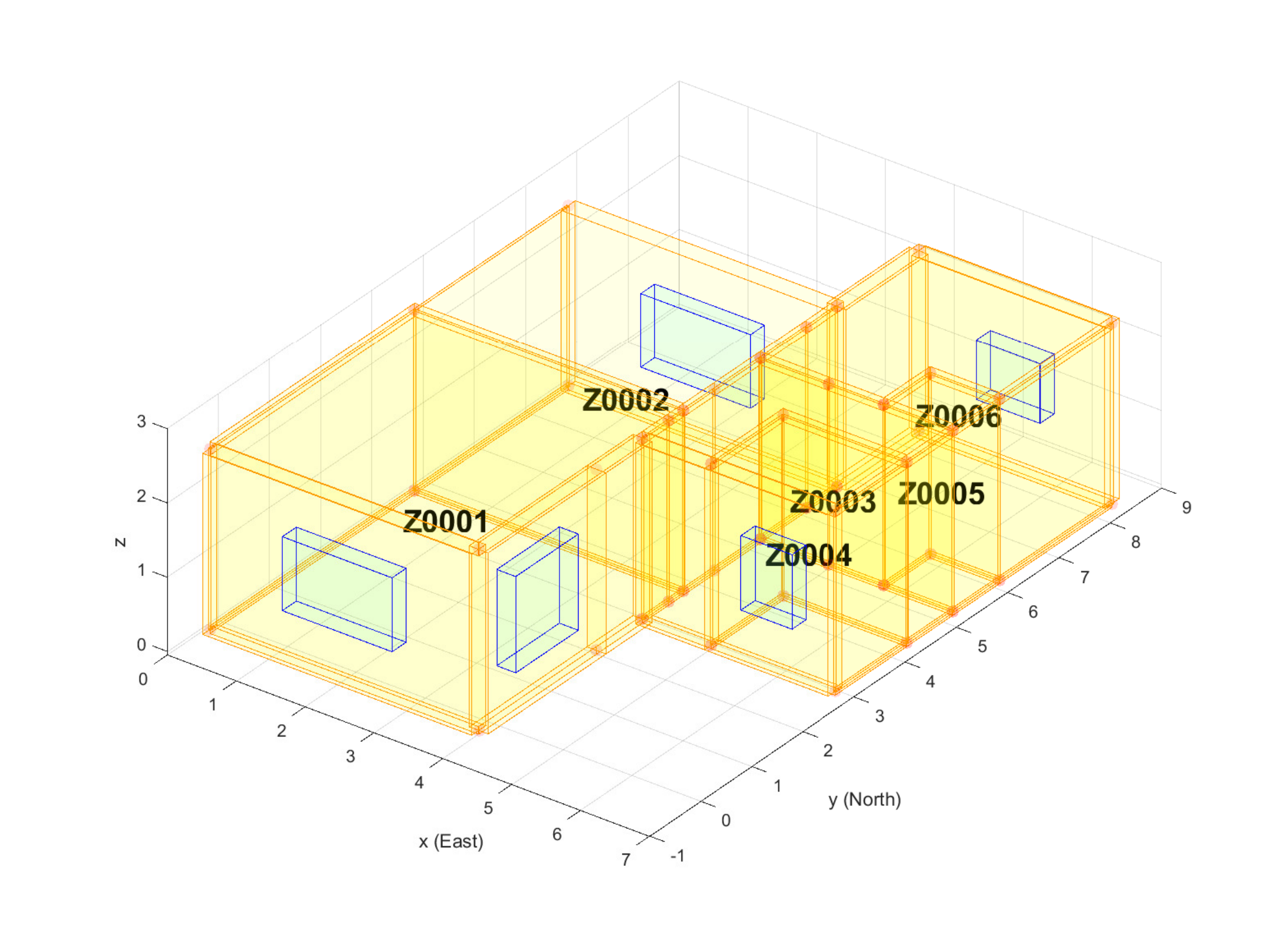}
\caption{Schematic of six-room apartment used for the building energy case study}
\label{fig_flat}
\end{figure}

A popular technique for building thermal simulation is to represent the structure as a Resistance Capacitance (RC) network~\cite{ODwyer2016}, particularly when knowledge of the physical composition of the building is available. The materials making up the walls, floors, ceilings, and windows are represented as configurations of resistances and capacitances whereby current flows through the circuit are analogous to heat flows through building components. 
Here, an EnergyPlus model of a six-room apartment was created based on standard building materials and thermal behaviour characteristics taken from the Tabula Webtool~\cite{Tabula}, and the underlying thermal model was extracted using the Building Resistance-Capacitance Modelling (BRCM) toolbox~\cite{Sturzenegger2014}. This resulting thermal model can be represented as a 102-state, linear, state-space system with six inputs (radiators in each room) and six outputs (the room temperatures). A schematic of the apartment layout can be seen in Fig.~\ref{fig_flat}.

The building model is influenced by the ambient temperature and solar irradiance from different orientations. For this, weather data from a London-based weather station was obtained from the CEDA archive~\cite{MET2006}. The occupancy profile used in the simulation was taken from the occupancy-integrated archetype approach of~\cite{Buttitta2019}. During occupied periods, a comfort set-point band between 20$^\circ$C and 22$^\circ$C was desired, while in unoccupied times, the temperatures were allowed to vary between 16$^\circ$C and 26$^\circ$C. The input in each room was constrained between 0 and the upper heat supply limit of the radiator in the room. The radiators were sized to emit a maximum of 100W per m$^{2}$ of floor area. The simulation ran with a 10-second sample time.

A separate data-driven predictive controller in each room with a sample time of 15 minutes was used to dictate the heat flow from the radiator to the room. The future set-point requirements were known to the controllers, as well as the current and previous room temperature and heat flow measurements. No measurements or forecasts of the weather were available to the controllers. A training period was carried out in which the radiators attempted to track a set-point varying between the upper and lower set-point bounds, using a PI controller. Note that this approach implies that the comfort set-point bounds should not be violated during the training period. The length of the training period depended on the formulation used (segmented or unsegmented) and the prediction horizon chosen for a particular scenario.

A set of scenarios were designed to compare the performance of the segmented and unsegmented formulations using different prediction horizons in this simulation environment. For these scenarios, we seek to minimise the deviation of the room temperatures outside the comfort bounds at a minimal cost. It was assumed that a heat pump supplies heat to the radiators with a Coefficient of Performance (COP) of 2.5, with electricity purchased via a time-varying tariff. For this, wholesale electricity price data was used with the Octopus Agile pricing tariff mechanism applied~\cite{Octopus}.

The formulation of Section~\ref{sec:example1} was modified slightly to incorporate an energy cost in the objectives. Once again, a prioritised framework was used, with the slack variables minimised first, followed by discomfort minimisation in a second optimisation, before finally minimising energy cost.
The first two optimisation levels are formulated as in (\ref{eq:obj1})--(\ref{eq:objend}). The financial cost is considered in the third optimisation problem. The predicted electricity price for the period from~$k+1$ to~$k+N$ is given as~$C_{elec}=(c[k+1],\ldots,c[k+N])\in\mathbb{R}^{N}$. The predicted electricity cost for heat pump consumption associated with the room over the prediction horizon was then included in the third-level objective as follows:
\begin{equation}
    \min_{\scriptsize{\begin{array}{c}g_{1}\ldots,g_{F},\varepsilon_{y},\\\varepsilon_{f_{1}},\ldots,\varepsilon_{f_{F}}\end{array}}}\eta C_{elec}I_{F}\otimes U_{\beta_{s}}\left[\begin{matrix}g_{1}\\\vdots\\g_{F}\end{matrix}\right]+\lambda_{g}\sum_{i=1}^{F}g_{i}^{T}g_{i},
\end{equation}
where~$\eta$ denotes the heat pump COP.

The constraints for this third-level problem are the same as for the second-level problem, with an additional constraint needed to enforce the optimal comfort performance, given as
\begin{equation}
    ||\varepsilon_{y}||_{1}\leq ||\varepsilon_{y}^{*}||_{1},
\end{equation}
where~$\varepsilon_{y}^{*}$ is the optimal~$\varepsilon_{y}$ computed in the second-level optimisation problem. A decentralised architecture was used, in which each room has a separate controller and no communication between controllers occurs.

\subsection{Performance analysis of data-driven controllers}
\begin{figure}[!tb]
\centering
\includegraphics[width=\mysize\textwidth]{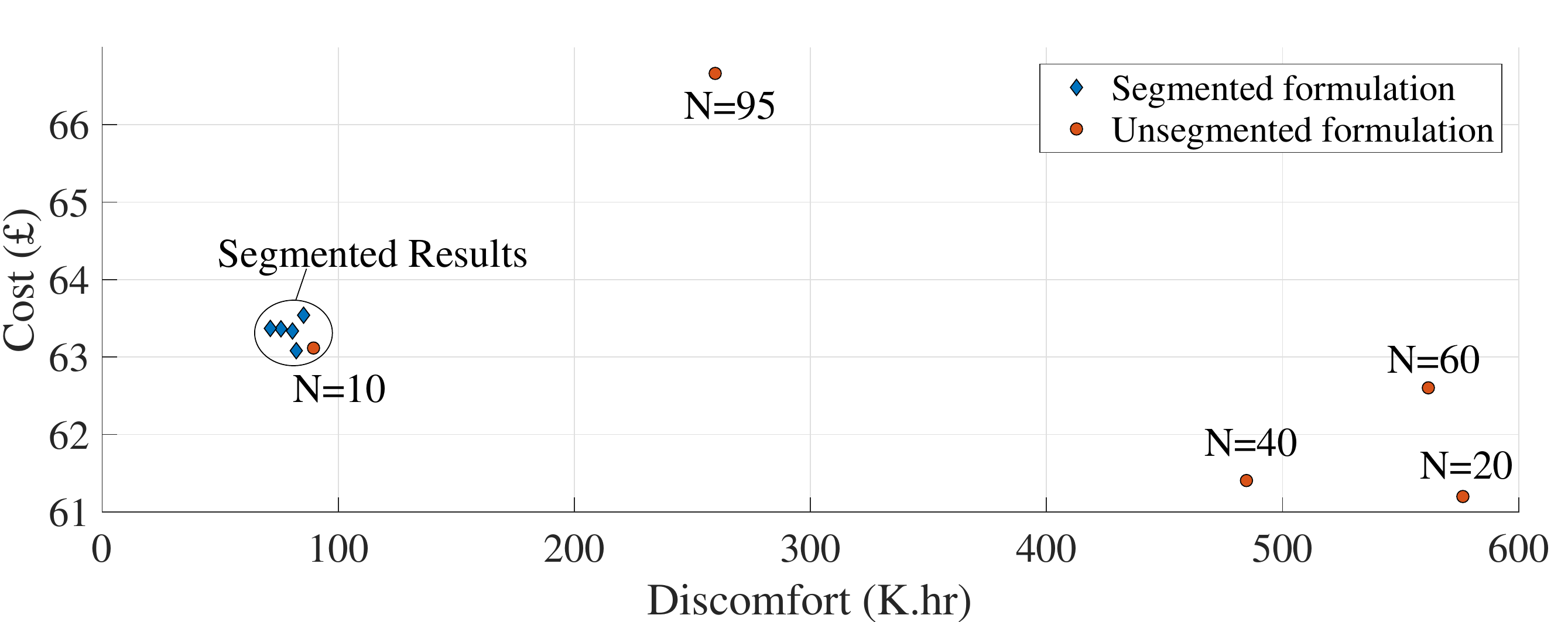}
\caption{Cost versus comfort objectives for different prediction horizons using segmented and unsegmented formulations}
\label{fig_scatt}
\end{figure}

Simulations were carried out to analyse the performance of the controllers for a three-week period using different prediction horizon lengths with the segmented and unsegmented formulations. Prediction horizons from 10 samples (2.5 hours) to 95 samples (just under 1 day) are investigated. In all cases, $T_{ini}=5$ and~$\lambda_{g}=1$ as these values were found to perform best for both segmented and unsegmented formulations. The results are summarised in Fig.~\ref{fig_scatt}, where the total heating cost for the apartment is plotted on the Y-axis and a discomfort metric is plotted on the X-axis. This discomfort metric is defined as the summation of absolute deviations from the comfort temperature set-point band in Kelvin (K), summed across each zone, scaled appropriately to achieve units of K$\cdot$hr.
 
The first noticeable feature of the results is that the segmented results for all horizon choices are closely grouped together, whereas the unsegmented results vary quite a lot in terms of both cost and comfort. Furthermore, in terms of comfort, all unsegmented cases with prediction horizons longer than $2.5$ hours fare significantly worse than the segmented cases. Although the unsegmented formulation achieves a lower cost than the segmented formulation for horizons longer than 2.5 hours, this comes at the expense of more discomfort. Since the objectives are prioritised and the comfort objective has a higher priority than the financial one, such behaviour is not indicative of a trade-off between objectives, it is indicative of poor control performance of the unsegmented formulation. 

Table~\ref{table_res1} provides the underlying values of the discomfort results. An interesting trend that can be seen in the segmented formulation is that longer prediction horizons lead to improved comfort. This is expected since pre-heating can be better exploited with longer predictions. Without segmentation, this benefit is not realised as control performance is compromised with longer predictions. 

\begin{table}[!tb]
\renewcommand{\arraystretch}{1.3}
\caption{Discomfort metric for unsegmented and segmented formulations with different horizons (3-week simulation)}
\label{table_res1}
\centering
    \begin{tabular}{M{1.5cm}||M{.9cm}|M{.9cm}|M{.9cm}|M{.9cm}|M{.9cm}}
     &\resizebox{0.9cm}{!}{$N=10$}&\resizebox{0.9cm}{!}{$N=20$}&\resizebox{0.9cm}{!}{$N=40$}&\resizebox{0.9cm}{!}{$N=60$}&\resizebox{0.9cm}{!}{$N=95$}\\
    \hline
    \hline
    Unsegmented (K$\cdot$hr) & 89.5 & 576.4 & 484.7 & 561.7 & 259.7\\
    \hline
    Segmented (K$\cdot$hr) & 85.3 & 82.2 & 80.6 & 75.7 & 71.2\\
    \hline
    \end{tabular}
\end{table}

The heating cost associated with each scenario is provided in Table~\ref{table_res2}. The segmented costs are similar for all horizon choices. For the unsegmented cases, low costs are achieved in some cases, for example, $N=20$ and $N=40$, however, these costs come at the expense of comfort. Only the $N=10$ unsegmented formulation is competitive with the segmented formulation.

\begin{table}[!tb]
\renewcommand{\arraystretch}{1.3}
\caption{Heating cost for unsegmented and segmented formulations with different horizons (3-week simulation)}
\label{table_res2}
\centering
    \begin{tabular}{M{1.5cm}||M{.9cm}|M{.9cm}|M{.9cm}|M{.9cm}|M{.9cm}}
     &\resizebox{0.9cm}{!}{$N=10$}&\resizebox{0.9cm}{!}{$N=20$}&\resizebox{0.9cm}{!}{$N=40$}&\resizebox{0.9cm}{!}{$N=60$}&\resizebox{0.9cm}{!}{$N=95$}\\
    \hline
    \hline
    Unsegmented (\pounds) & 63.1 & 61.2 & 61.4 & 62.6 & 66.7\\
    \hline
    Segmented (\pounds) & 63.5 & 63.1 & 63.3 & 63.4 & 63.4\\
    \hline
    \end{tabular}
\end{table}

\begin{figure}[!tb]
    \centering
    \includegraphics[width=\mysize\textwidth]{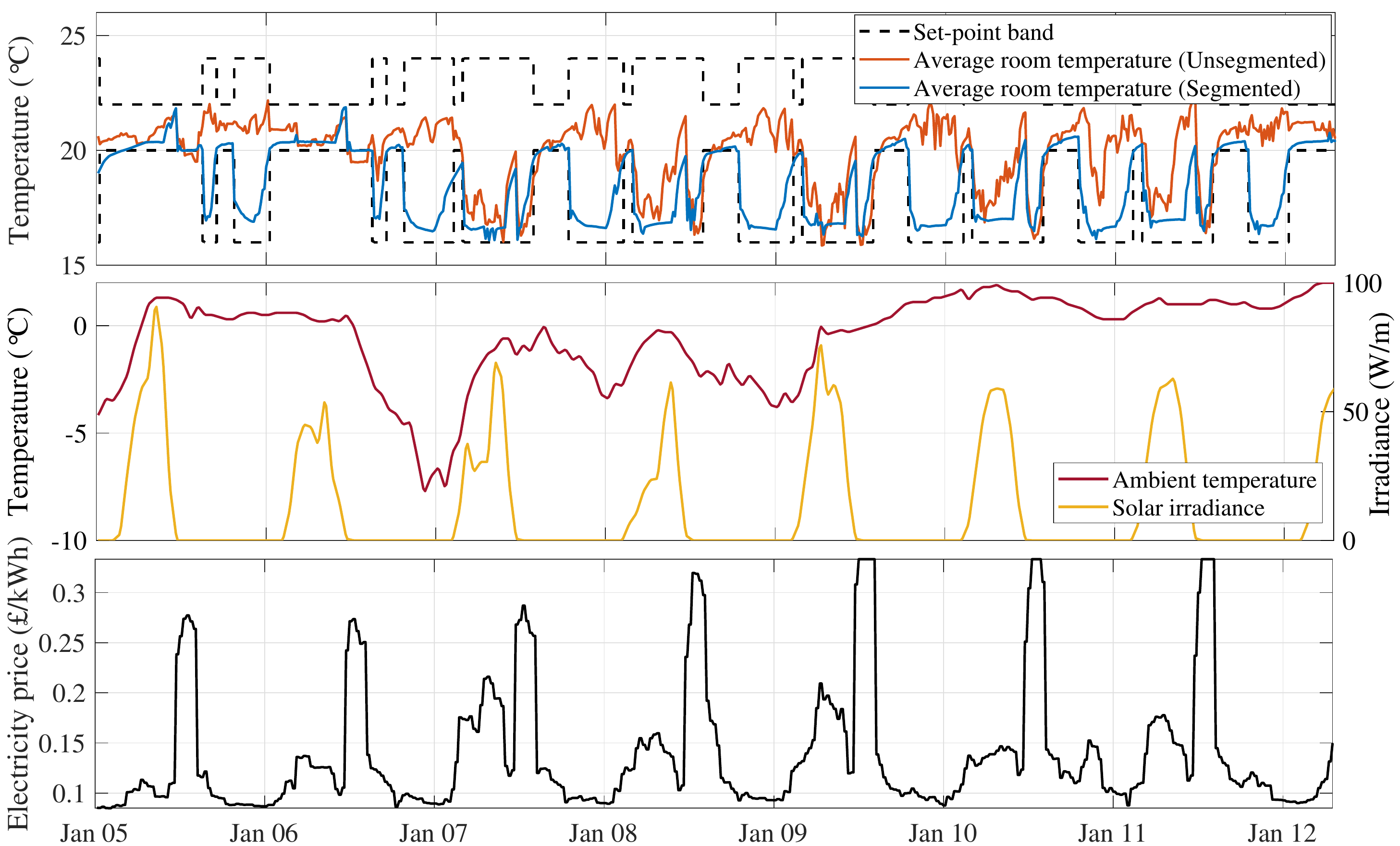}
    \caption{Average room temperature in building with segmented and unsegmented formulations, plotted for one week of the simulation period ($N=95$), and corresponding external weather conditions and electricity price profile for the period}
    \label{fig_AvTemp}
\end{figure}

A one-week window of the average apartment temperatures using~$N=95$ with the segmented and unsegmented formulations is plotted in Fig.~\ref{fig_AvTemp} to illustrate the differing control performance. The electricity price for the same period is also shown. The unsegmented formulation overheats the apartment during the unoccupied periods compared to the segmented formulation. In both formulations, the controllers tend to pre-heat the apartment in advance of an electricity price spike, however, without segmentation, the rooms are held at a higher temperature than is necessary.

As in the examples of Section~\ref{sec:example1}, the performance of the unsegmented strategy breaks down as the prediction horizon length increases, while the segmented formulation is more consistent in a wider range of operational strategies.

\section{Conclusions}\label{sec:End}
This paper proposes a restructuring of a data-driven predictive control formulation for linear systems with unmeasured disturbances and noise. The proposed formulation modifies an existing data-enabled predictive control approach by segmenting the prediction horizon. By doing so, the formulation performs better than the unsegmented formulation in the presence of unmeasured disturbance, particularly for longer prediction horizons.

The method was analysed here first using a set of case studies based on a two-mass-spring-damper system. Under various disturbance and noise realisations, the segmented formulation outperformed the unsegmented formulation in terms of set-point tracking when disturbances were present, particularly with longer prediction horizons. The computation time associated with the proposed segmented formulation scales linearly with horizon length, improving on the time increase observed for the unsegmented formulation.

The segmented formulation was applied to a building energy management case study to demonstrate the importance of these performance characteristics in a more realistic setting, using a state-of-the-art building simulation environment with realistic weather profiles acting as unmeasured disturbances. The segmented formulation performed consistently with horizon length variation in terms of occupant comfort levels and energy consumption. Without segmentation, the comfort minimisation performance of the controller was significantly worse for prediction horizons longer than 10 samples. For the scenario with a one-day-ahead prediction horizon, the segmented approach reduced discomfort by 72\% and cost by 5\% relative to the unsegmented approach. 

Further work is needed to assess the impact of segmentation on the various extensions of the data-predictive controller that have been developed, such as formulations with robustness guarantees and formulations for time-varying parameters and nonlinear systems. Additionally, methods for offset-free control in the presence of disturbance would also be beneficial to the data-driven context. Computational efficiency and hyperparameter selection are also key aspects that require further focus to ensure algorithms are tailored appropriately to a given context.

\section*{Acknowledgement}

This work has received funding from the EPSRC (Engineering and Physical Sciences) under the Active Building Centre project (reference number: EP/V012053/1).

\bibliographystyle{IEEEtran}
\bibliography{bibfile.bib}

\begin{thebibliography}{10}
\providecommand{\url}[1]{#1}
\csname url@samestyle\endcsname
\providecommand{\newblock}{\relax}
\providecommand{\bibinfo}[2]{#2}
\providecommand{\BIBentrySTDinterwordspacing}{\spaceskip=0pt\relax}
\providecommand{\BIBentryALTinterwordstretchfactor}{4}
\providecommand{\BIBentryALTinterwordspacing}{\spaceskip=\fontdimen2\font plus
\BIBentryALTinterwordstretchfactor\fontdimen3\font minus
  \fontdimen4\font\relax}
\providecommand{\BIBforeignlanguage}[2]{{%
\expandafter\ifx\csname l@#1\endcsname\relax
\typeout{** WARNING: IEEEtran.bst: No hyphenation pattern has been}%
\typeout{** loaded for the language `#1'. Using the pattern for}%
\typeout{** the default language instead.}%
\else
\language=\csname l@#1\endcsname
\fi
#2}}
\providecommand{\BIBdecl}{\relax}
\BIBdecl

\bibitem{Maddalena2020a}
E.~T. Maddalena, Y.~Lian, and C.~N. Jones, ``{Data-driven methods for building
  control - A review and promising future directions},'' \emph{Control
  Engineering Practice}, vol.~95, no. June 2019, p. 104211, 2020.

\bibitem{Bhattacharyya2021}
V.~Bhattacharyya, A.~F. Canosa, and B.~HomChaudhuri, ``{Fast Data-Driven Model
  Predictive Control Strategy for Connected and Automated Vehicles},''
  \emph{ASME Letters in Dynamic Systems and Control}, vol.~1, no.~4, pp. 1--5,
  2021.

\bibitem{Wang2019b}
J.~Wang, S.~Li, H.~Chen, Y.~Yuan, and Y.~Huang, ``{Data-driven model predictive
  control for building climate control: Three case studies on different
  buildings},'' \emph{Building and Environment}, vol. 160, no. March, p.
  106204, 2019.

\bibitem{DePersis2020}
C.~{De Persis} and P.~Tesi, ``{Formulas for Data-Driven Control: Stabilization,
  Optimality, and Robustness},'' \emph{IEEE Transactions on Automatic Control},
  vol.~65, no.~3, pp. 909--924, 2020.

\bibitem{Willems2005}
J.~C. Willems, P.~Rapisarda, I.~Markovsky, and B.~L. {De Moor}, ``{A note on
  persistency of excitation},'' \emph{Systems and Control Letters}, vol.~54,
  no.~4, pp. 325--329, 2005.

\bibitem{Coulson2019a}
J.~Coulson, J.~Lygeros, and F.~Dorfler, ``{Data-enabled predictive control: In
  the shallows of the {DeePC}},'' \emph{2019 18th European Control Conference,
  ECC 2019}, pp. 307--312, 2019.

\bibitem{Fiedler2020}
\BIBentryALTinterwordspacing
F.~Fiedler and S.~Lucia, ``{On the relationship between data-enabled predictive
  control and subspace predictive control},'' 2020. [Online]. Available:
  \url{http://arxiv.org/abs/2011.13868}
\BIBentrySTDinterwordspacing

\bibitem{Dorfler2021}
\BIBentryALTinterwordspacing
F.~D{\"{o}}rfler, J.~Coulson, and I.~Markovsky, ``{Bridging direct {\&}
  indirect data-driven control formulations via regularizations and
  relaxations},'' 2021. [Online]. Available:
  \url{http://arxiv.org/abs/2101.01273}
\BIBentrySTDinterwordspacing

\bibitem{VanWaarde2020}
H.~J. {Van Waarde}, C.~{De Persis}, M.~K. Camlibel, and P.~Tesi, ``{Willems'
  Fundamental Lemma for State-Space Systems and Its Extension to Multiple
  Datasets},'' \emph{IEEE Control Systems Letters}, vol.~4, no.~3, pp.
  602--607, 2020.

\bibitem{Alpago2020}
D.~Alpago, F.~Dorfler, and J.~Lygeros, ``{An Extended Kalman Filter for
  Data-Enabled Predictive Control},'' \emph{IEEE Control Systems Letters},
  vol.~4, no.~4, pp. 994--999, 2020.

\bibitem{Berberich2021}
J.~Berberich, J.~Kohler, M.~A. Muller, and F.~Allgower, ``{Data-Driven Model
  Predictive Control with Stability and Robustness Guarantees},'' \emph{IEEE
  Transactions on Automatic Control}, vol.~66, no.~4, pp. 1702--1717, 2021.

\bibitem{Coulson2020}
\BIBentryALTinterwordspacing
J.~Coulson, J.~Lygeros, and F.~D{\"{o}}rfler, ``{Distributionally Robust Chance
  Constrained Data-enabled Predictive Control},'' no.~Id, pp. 1--14, 2020.
  [Online]. Available: \url{http://arxiv.org/abs/2006.01702}
\BIBentrySTDinterwordspacing

\bibitem{Huang2021}
\BIBentryALTinterwordspacing
L.~Huang, J.~Zhen, J.~Lygeros, and F.~D{\"{o}}rfler, ``{Robust Data-Enabled
  Predictive Control: Tractable Formulations and Performance Guarantees},''
  2021. [Online]. Available: \url{http://arxiv.org/abs/2105.07199}
\BIBentrySTDinterwordspacing

\bibitem{Lian2021}
\BIBentryALTinterwordspacing
Y.~Lian and C.~N. Jones, ``{From System Level Synthesis to Robust Closed-loop
  Data-enabled Predictive Control},'' 2021. [Online]. Available:
  \url{http://arxiv.org/abs/2102.06553}
\BIBentrySTDinterwordspacing

\bibitem{Lian2021b}
Y.~Lian, J.~Shi, M.~P. Koch, and C.~N. Jones, ``{Adaptive Robust Data-driven
  Building Control via Bi-level Reformulation: an Experimental Result},'' pp.
  1--12, 2021.

\bibitem{Willems2007}
J.~C. Willems, ``{The Behavioral Approach to Open and Interconnected
  Systems},'' \emph{IEEE Control Systems}, vol.~27, no.~6, pp. 46--99, 2007.

\bibitem{Markovsky2008}
I.~Markovsky and P.~Rapisarda, ``{Data-driven simulation and control},''
  \emph{International Journal of Control}, vol.~81, no.~12, pp. 1946--1959,
  2008.

\bibitem{Qin2005}
S.~J. Qin, W.~Lin, and L.~Ljung, ``{A novel subspace identification approach
  with enforced causal models},'' \emph{Automatica}, vol.~41, no.~12, pp.
  2043--2053, 2005.

\bibitem{Kerrigan2002}
E.~C. Kerrigan and J.~M. Maciejowski, ``{Designing model predictive controllers
  with prioritised constraints and objectives},'' \emph{IEEE International
  Symposium on Computer Aided Control System Design}, pp. 33--38, 2002.

\bibitem{Data_Chinde2022}
V.~Chinde, Y.~Lin, and M.~J. Ellis, ``Data-enabled predictive control for
  building hvac systems,'' \emph{Journal of Dynamic Systems, Measurement, and
  Control}, vol. 144, no.~8, p. 081001, 2022.

\bibitem{ODwyerEcos}
E.~O'Dywer, P.~Falugi, N.~Shah, and E.~Kerrigan, ``{Automating the data-driven
  predictive control design process for building thermal management},'' in
  \emph{35th International Conference on Efficiency, Cost, Optimization,
  Simulation and Environmental Impact of Energy Systems}, Copenhagen, 2022.

\bibitem{Lowes2020}
R.~Lowes, J.~Rosenow, M.~Qadrdan, and J.~Wu, ``{Hot stuff: Research and policy
  principles for heat decarbonisation through smart electrification},''
  \emph{Energy Research and Social Science}, vol.~70, no. March, p. 101735,
  2020.

\bibitem{Atam2016}
E.~Atam and L.~Helsen, ``{Control-Oriented Thermal Modeling of Multizone
  Buildings: Methods and Issues: Intelligent Control of a Building System},''
  \emph{IEEE Control Systems}, vol.~36, no.~3, pp. 86--111, 2016.

\bibitem{Kathirgamanathan2021}
A.~Kathirgamanathan, M.~{De Rosa}, E.~Mangina, and D.~P. Finn, ``{Data-driven
  predictive control for unlocking building energy flexibility: A review},''
  \emph{Renewable and Sustainable Energy Reviews}, vol. 135, no. January 2020,
  p. 110120, 2021.

\bibitem{DOE2017}
\BIBentryALTinterwordspacing
DOE, ``{EnergyPlus | EnergyPlus},'' p.~1, 2017. [Online]. Available:
  \url{https://www.energyplus.net/}
\BIBentrySTDinterwordspacing

\bibitem{ODwyer2016}
E.~O'Dwyer, L.~{De Tommasi}, K.~Kouramas, M.~Cychowski, and G.~Lightbody,
  ``{Modelling and disturbance estimation for model predictive control in
  building heating systems},'' \emph{Energy and Buildings}, vol. 130, pp.
  532--545, oct 2016.

\bibitem{Tabula}
\BIBentryALTinterwordspacing
TABULA, ``{TABULA webtool}.'' [Online]. Available:
  \url{http://webtool.building-typology.eu/{\#}bm}
\BIBentrySTDinterwordspacing

\bibitem{Sturzenegger2014}
D.~Sturzenegger, D.~Gyalistras, V.~Semeraro, M.~Morari, and R.~S. Smith,
  ``{BRCM Matlab Toolbox: Model generation for model predictive building
  control},'' \emph{Proceedings of the American Control Conference}, pp.
  1063--1069, 2014.

\bibitem{MET2006}
\BIBentryALTinterwordspacing
{Met Office (2006)}, ``{MIDAS}: {UK} hourly weather observation data. {NCAS
  British Atmospheric Data Centre}.'' [Online]. Available:
  \url{https://catalogue.ceda.ac.uk/uuid/916ac4bbc46f7685ae9a5e10451bae7c}
\BIBentrySTDinterwordspacing

\bibitem{Buttitta2019}
G.~Buttitta, W.~J. Turner, O.~Neu, and D.~P. Finn, ``{Development of
  occupancy-integrated archetypes: Use of data mining clustering techniques to
  embed occupant behaviour profiles in archetypes},'' \emph{Energy and
  Buildings}, vol. 198, pp. 84--99, 2019.

\bibitem{Octopus}
\BIBentryALTinterwordspacing
{Octopus Energy}, ``{Frequently asked questions about Octopus Tracker | Octopus
  Energy},'' 2020. [Online]. Available:
  \url{https://octopus.energy/tracker-faqs/}
\BIBentrySTDinterwordspacing

\end{thebibliography}

\end{document}